\documentclass[aps,prb,reprint,amsmath,amssymb,floatfix,superscriptaddress]{revtex4-2}

\usepackage{bm,dcolumn,graphicx,multirow}
\usepackage[caption=false,subrefformat=parens,labelformat=parens]{subfig}
\usepackage[colorlinks=true,citecolor=blue,urlcolor=black,bookmarks=false,
  hypertexnames=true]{hyperref}
\usepackage[ignoreunlbld,norefs,nocites]{refcheck}

\newcolumntype{d}[1]{D{.}{.}{#1}}

\usepackage[dvipsnames]{xcolor}

\begin{document}
% *{{ titles, authors, affiliations

\title{Point defect formation energies in graphene from diffusion
  quantum Monte Carlo and density functional theory}

\author{D.\ M.\ Thomas} \email{davidthomasuk0@gmail.com}

\affiliation{Department of Physics, Lancaster University, Lancaster
  LA1 4YB, United Kingdom}

\author{Y.\ Asiri}

\affiliation{Department of Physics, Lancaster University, Lancaster
  LA1 4YB, United Kingdom}

\affiliation{Department of Physics, King Khalid University, Abha, KSA}

\author{N.\ D.\ Drummond}

\affiliation{Department of Physics, Lancaster University, Lancaster
  LA1 4YB, United Kingdom}

\date{\today}

% }}*

% *{{ abstract, PACS, and maketitle
\begin{abstract}
Density functional theory (DFT) is widely used to study defects in
monolayer graphene with a view to applications ranging from water
filtration to electronics to investigation of radiation damage in
graphite moderators.
To assess the accuracy of DFT in such applications, we report
diffusion quantum Monte Carlo (DMC) calculations of the formation
energies of some common and important point defects in monolayer
graphene: monovacancies, Stone-Wales defects, and silicon
substitutions.
We find that standard DFT methods underestimate monovacancy formation
energies by around $1$ eV\@.
The disagreement between DFT and DMC is somewhat smaller for
Stone-Wales defects and silicon substitutions.
We examine vibrational contributions to the free energies of formation
for these defects, finding that vibrational effects are
non-negligible.
Finally, we compare the DMC atomization energies of monolayer
graphene, monolayer silicene, and bulk silicon, finding that bulk
silicon is significantly more stable than monolayer silicene by
$0.7522(5)$ eV per atom.
\end{abstract}

%\keywords{quantum Monte Carlo}
\maketitle

% }}*

%\tableofcontents

% *{{ Introduction

\section{Introduction}

Graphene, an atomically thin sheet of carbon atoms forming a honeycomb
lattice, is one of the most promising materials for future
technological applications
\cite{Novoselov2004,Castro_2009,Sarma_2011}.
However, producing large, defect-free sheets of graphene on insulating
substrates remains a significant technological challenge
\cite{Munoz_2013}.
Point defects may appear naturally during the growth of graphene, or
they may be deliberately inserted into pristine graphene by processing
\cite{Vicarelli_2015}.
Point defects can have a major impact on the electronic and optical
properties of graphene
\cite{huang_meyer_muller_2012,zhang2016opening}, so it is necessary to
understand their structure and characteristics to gain a full
understanding of the performance of graphene-based devices.
High-resolution transmission electron microscopy and related
techniques have been employed to obtain clear imaging of defect
structures in graphene \cite{Hashimoto_2004,Robertson_2013},
but these methods inevitably introduce further defects.
Theoretical methods have also played a key role in studies of defects
in graphene.
In particular, there are numerous works in which density functional
theory (DFT) has been used to evaluate defect formation energies and other
properties pertaining to a range of applications and devices featuring
graphene \cite{Lusk_2008,Yang_2015,Okamoto_2016}, graphite
\cite{Kaxiras_1988,ElBarbary_2003,Li_2005}, and other two-dimensional
(2D) or layered materials
\cite{Azevedo_2007,Reimers_2018,Blades_2020}.
The main purpose of the present work is to provide quantum Monte Carlo
(QMC) defect-formation energy data to assess the accuracy of DFT in
studies of defects in graphene.

Monovacancies (MVs) in graphene have been studied for both their
desired and undesired effects on the graphene lattice.
Graphite has long been used as a neutron moderator in nuclear
reactors, which exposes the lattice to radiation damage
\cite{Banhart1991,Hahn1999,Telling_2007}.
It is essential to understand the formation of radiation-induced
vacancies and how these defects alter or weaken the structure of
graphene layers, and in turn graphite itself \cite{ElBarbary_2003}.
Vacancies in graphene also arise due to damage by electron beams in
transmission electron microscopy \cite{Bachmatiuk_2015}.
Vacancy defects can in fact be useful for some applications and hence
may be deliberately introduced into the lattice.
Graphite/graphene has commonly been used as an anode material in
lithium-ion batteries, with the lithium ions able to intercalate in
the lattice \cite{Goriparti2014}.
A move towards sodium- or calcium-ion batteries is desirable owing to
the greater abundance and lower cost of these metals; unfortunately,
the larger size of sodium and calcium ions inhibits intercalation.
However, the additional space created by vacancy defects allows larger
atoms to intercalate into the anode material
\cite{FarokhNiaei2018,Palomares2012}.
Likewise, sub-nm pores, of which the MV is the smallest possible
example, allow ion-selective transport for applications such as
desalination of seawater \cite{Lee_2014,Han_2018}.
The studies cited here depend on DFT calculations to explore the
behavior of MVs and their interaction with other defects and chemical
species.

The most important feature in the electronic structure of graphene is
the Dirac point at the Fermi level of pristine graphene.
Many types of defect at finite concentration break the sublattice
symmetry and/or shift the Fermi level, significantly altering the
electronic properties of graphene \cite{banhart2010structural}.
Substitutional impurity atoms are among the most common defects in
graphene, and have been extensively studied using DFT
\cite{Wadey2016}.
Several studies
\cite{ci2010atomic,lazar2014chemical,rani2014stability} have shown
that nitrogen and boron impurities in graphene act as donors and
acceptors, respectively.
DFT has been used to investigate the electronic and magnetic
properties of a graphene sheet doped by iron, cobalt, silicon, and
germanium impurities at $3$\% concentration, finding that the
substitution of a carbon atom with silicon or germanium can open a
band gap in the electronic spectrum of graphene, while the insertion
of iron or cobalt produces a metallic phase
\cite{kheyri2016electronic}.
Silicon substitutions (SiSs) in graphene are an attractive approach
for engineering the band structure \cite{ervasti2015silicon}.
The silicon atom, which has the same number of valence electrons as
carbon, has been shown to be able to modulate the electronic structure
of graphene without significantly changing its carrier mobility
\cite{zhang2016opening}.

Stone-Wales (SW) defects in graphene are some of the most commonly
observed intrinsic topological defects \cite{banhart2010structural}.
SW defects influence the electronic, structural, chemical, and
mechanical properties of graphene
\cite{kang2008effect,podlivaev2015out,carlsson2006structural,
  boukhvalov2008chemical,juneja2018anomalous,sagar2020effect}.
SW defects result in a tendency for monolayer graphene to bend, and
therefore can be used in the fabrication of nonplanar carbon
nanostructures \cite{openov2015interaction}.
SW defects show mutual attraction \cite{Podlivaev_2015}, and the
formation of clusters of SW defects at high temperature is one of the
first steps in the melting of graphene \cite{Zakharchenko_2011}.
Once again, DFT has played a key role in elucidating the properties of
SW defects.

The single most important thermodynamic property of a point defect is
its formation energy $\mathcal{E}^\text{f}$, which is the difference
in free energy between the defected material and the pristine
material, together with any changes in the energies of reservoirs of
the atoms that are added or removed when the defect is formed (i.e.,
$\mathcal{E}^\text{f}$ is the change in the grand potential when the
defect is formed).
At zero external stress and low defect concentration, and assuming
thermal equilibrium with appropriate reservoirs, the point defect
concentration in a 2D material is given by
$n=\exp[-\mathcal{E}^\text{f}/(k_\text{B}T)]/A$, where $k_\text{B}$ is
Boltzmann's constant, $T$ is temperature, and $A$ is the
primitive-cell area.

In this paper, we present QMC calculations of the formation energies
of isolated MVs, SiSs, and SW defects.
Our intention is to benchmark the accuracy of the DFT methods that
have been widely used in studies of defects in graphene.
Given the lack of precise experimental results in this area,
comparing first-principles methods in this way is the only method of
assessing the accuracy of the calculations.
Since this work necessitates the calculation of the energy per atom of
both graphene and bulk silicon, we also take this opportunity to
investigate the energetic stability of silicene, the 2D allotrope of
silicon, relative to bulk diamond-structure silicon.

Silicene, the silicon counterpart of graphene, is a honeycomb
structure of silicon atoms with slightly buckled hexagonal sublattices
that result from mixing $sp^2$ and $sp^3$ hybridization.
The dynamical stability of free-standing silicene has been
demonstrated theoretically using DFT calculations
\cite{roome2014beyond,cahangirov2009two}; however, in practice it can
only be synthesized experimentally on metal surfaces
\cite{vogt2012silicene,cinquanta2013getting,meng2013buckled}.
This new 2D material has been extensively studied theoretically due to
the many remarkable properties that result from its relatively large
spin-orbit coupling and buckled structure
\cite{drummond2012electrically,kaloni2016current}.
Under external electric field, silicene undergoes a transition from a
topologically insulating phase arising from the spin-orbit coupling
into a variety of quantum phases
\cite{ezawa2013spin,bao2017photoinduced}.
Silicene holds great promise for a variety of applications in
spintronic and optoelectronic devices
\cite{chowdhury2016theoretical,molle2018silicene}.
However, a fundamental issue for any attempt to use silicene in
practical devices is its lack of thermodynamic stability.
Here, we report QMC simulations performed to compare the ground-state
energies per atom of bulk silicon and free-standing silicene.

Our calculations make use of two QMC methods, namely variational Monte
Carlo (VMC) and diffusion Monte Carlo (DMC), in tandem.
VMC uses Monte Carlo integration to evaluate expectation values with
respect to a many-body trial wave function; this, in combination with
the variational principle, can be used to optimize free parameters in
the trial wave function.
The resulting optimized wave function is then used as the starting
point for a DMC calculation.
The DMC method projects out the ground-state component of the wave
function by simulating the evolution of a population of walkers
governed by the imaginary-time-dependent Schr\"{o}dinger equation.
Fermionic antisymmetry is maintained using the fixed-phase
approximation \cite{Anderson1976,Ortiz_1993}.
All our QMC calculations were performed using the \textsc{casino} code
\cite{Needs_2020} to study supercells of graphene, bulk silicon, and
silicene subject to twisted periodic boundary conditions.
A ``twist-blocking'' method is introduced to evaluate the error on
twist-averaged results, accounting for the random errors due to both
Monte Carlo integration and the random sampling of twists.

The rest of the paper is organized as follows.
In Sec.\ \ref{sec:method} we describe our computational methodology.
In Sec.\ \ref{sec:results} we present and analyze our numerical
results.
Finally we draw our conclusions in Sec.\ \ref{sec:conc}.

% }}*

% *{{ Computational methodology

\section{Computational methodology}
\label{sec:method}

\subsection{Defect formation energies}

We define the ``pure'' formation energy $\mathcal{E}^\text{pf}$ of an isolated
defect in graphene as the free-energy difference between a large
region of graphene containing a single defect and pristine graphene.
The defect formation energy $\mathcal{E}^\text{f}$ is the sum of the
pure defect formation energy and the changes in the free energies of
reservoirs of the atoms that are added or removed, i.e.,
$\mathcal{E}^\text{f}$ is defined via a difference in grand
potentials.
For the SW defect, MV, and SiS these are
\begin{eqnarray}
\mathcal{E}^\text{f}_\text{MV}&=&\mathcal{E}^\text{pf}_\text{MV}+\mu_\text{C}
\label{eq:MV}
\\ \mathcal{E}^\text{f}_\text{SiS}&=&\mathcal{E}^\text{pf}_\text{SiS}+\mu_\text{C}-\mu_\text{Si}
\label{eq:Si}
\\ \mathcal{E}^\text{f}_\text{SW}&=&\mathcal{E}^\text{pf}_\text{SW}, \label{eq:SW}
\end{eqnarray}
where the chemical potentials $\mu_\text{C}$ and $\mu_\text{Si}$ are
taken to be the Helmholtz free energies per atom of monolayer graphene
and bulk diamond-structure silicon, respectively.
The pure defect formation energy is not in general physically
meaningful by itself, because it depends on the choice of
pseudopotentials.
However, it is theoretically useful because it allows us to
distinguish finite-concentration and finite-size effects purely due to
defect formation in a periodic supercell from finite-size errors in
the energy per atom of graphene and silicon.
We approximate that the pure defect formation energy is the sum of the
difference of static-nucleus electronic ground-state energies of
defective and pristine graphene, which we evaluate by both DFT and
DMC, and the temperature-dependent difference of vibrational Helmholtz
free energies, which we evaluate by DFT\@.
Likewise, each chemical potential is taken to be the sum of the
static-nucleus electronic ground-state energy per atom and the
DFT-calculated temperature-dependent vibrational Helmholtz free energy
per atom.

Details of the DFT and DMC calculations can found in Appendix
\ref{app:comp_details}.

\subsection{Free energies of atomization}

We define the free energy of atomization of bulk silicon as the
difference of the energy of an isolated, spin-polarized silicon atom
in its $^3\text{P}_0$ ground state and the Helmholtz free energy per
atom in bulk silicon.
The atomization energies of silicene and graphene are defined in an
analogous manner.
This provides a pseudopotential-independent (in principle) free energy
per atom that can be used to compare the stability of different
condensed phases.
Note, however, that the temperature dependence of the free energy of
the reference gaseous atomic state is neglected.

% *{{ Finite-size effects

\subsection{Finite-concentration and finite-size effects}

\subsubsection{Periodic supercells}

Our QMC calculations of defect formation energies were performed in
finite supercells subject to periodic boundary conditions, with a
single point defect in the simulation cell.
This leads to a number of physical differences from the dilute limit
of isolated point defects in which we are interested.
Firstly, there are finite-concentration effects due to the fact that
we are simulating a periodic array of point defects rather than an
isolated defect.
Leading-order systematic finite-concentration effects are due to
screened electrostatic interactions between periodic images of defects
and elastic interactions between defects \cite{Castleton_2004}.
Finite concentration effects can also arise due to the unwanted
dispersion of localized defect states.
There are also nonsystematic finite-concentration effects due to
interactions between charge-density oscillations around defects.
We reduce the systematic effects and average out the nonsystematic
effects by extrapolation to infinite cell size using an appropriate
fitting function.
Secondly, at a given defect concentration there are finite-size
effects arising from the simulation of periodic supercells rather than
infinite systems.
These include quasirandom, oscillatory effects due to momentum
quantization, which we address by averaging over twisted boundary
conditions on the supercell \cite{Lin2001}.
Long-range finite-size effects largely cancel out of the pure defect
formation energies: the expressions for the leading-order corrections
are the same for pristine and defective cells.

To calculate chemical potentials we must find the ground-state
energies per atom of graphene and bulk silicon.
In a finite simulation cell these suffer from quasirandom
momentum-quantization effects as well as long-range effects due to the
evaluation of the interaction between each electron and the
surrounding exchange-correlation hole using the Ewald interaction
rather than $1/r$ \cite{Fraser_1996} and the neglect of long-range
two-body correlations \cite{Chiesa_2006,Drummond_2008}.

\subsubsection{Twist averaging}

Unlike DFT, only a single ${\bf k}$ point can be used in each QMC
calculation.
We use twist averaging in the canonical ensemble \cite{Lin2001} to
reduce momentum-quantization errors in our results.
All our graphene and bulk silicon DMC calculations were carried out at
$24$ random twists, while our silicene calculations used $48$ random
twists.
Since momentum quantization is a single-particle effect, it is well
described by DFT, so that the QMC and DFT energies are correlated as a
function of twist.
DFT energies can therefore be used as a control variate (CV) when
evaluating the twist-averaged (TA) DMC energy.
The TA energy $E_\text{DMC}^\text{TA}$ was found by fitting
\begin{equation} E_\text{DMC}({\bf k}_\text{s})=E_\text{DMC}^\text{TA}
+b\left[E_\text{DFT}({\bf k}_\text{s})-E_\text{DFT}^\text{fine}\right]
\label{eq:tafit} \end{equation}
to the DMC energy $E_\text{DMC}({\bf k}_\text{s})$ at twist ${\bf
  k}_\text{s}$, where $b$ is a fitting parameter and
$E_\text{DFT}({\bf k}_\text{s})$ is the corresponding DFT energy, and
$E_\text{DFT}^\text{fine}$ is the DFT energy evaluated using a fine
${\bf k}$-point grid.
Equation (\ref{eq:tafit}) simultaneously removes most of the
quasirandom noise due to momentum quantization and corrects for
residual errors in the TA energy by virtue of the fact that the
correlator is the DFT energy relative to the DFT energy with a fine
${\bf k}$-point mesh rather than the TA-DFT energy.
The pristine and defective graphene calculations were performed at
identical twists, so the twist-sampling error in the difference is
much smaller than the twist-sampling errors in the total energies.
When calculating the TA pure defect formation energy, we used DMC and
DFT pure defect formation energies $\mathcal{E}^\text{pf}({\bf
  k}_\text{s})$ in Eq.\ (\ref{eq:tafit}) rather than total energies
$E({\bf k}_\text{s})$.

There are two very different sources of (quasi-)random error in the
TA-DMC energy for a given supercell: the statistical error from the
Monte Carlo simulation, and the residual momentum quantization error
that is not fully removed by fitting Eq.\ (\ref{eq:tafit}) to
$\mathcal{E}^\text{pf}({\bf k}_\text{s})$.
The statistical error can easily be accounted for by Gaussian
propagation of errors; however the residual momentum quantization
error is unknown at the outset.
To quantify the remaining error, a ``twist-blocking'' procedure has
been used.
For example, $24$ twists can be grouped into four blocks of six
twists, and within each block the TA energy can be calculated by
fitting Eq.\ (\ref{eq:tafit}).
An estimate of the true TA energy is then given by the mean of the
four independent values of $E_{\rm DMC}^{\rm TA}$, while the standard
error in the mean quantifies both the residual momentum quantization
error and the Monte Carlo errors.
The mean energy obtained by this procedure is a biased estimate of the
TA energy due to the small number of twists used in each fit; however,
we can check for bias in both the mean and the standard error in the
mean by increasing the block size.
In fact we minimize the bias in the mean by using
Eq.\ (\ref{eq:tafit}) with all the twists to obtain the TA energy, and
only use the twist-blocking method to estimate the error bar on the TA
energy.

\begin{figure}[!htbp]
\centering
\includegraphics[scale=0.3,clip]{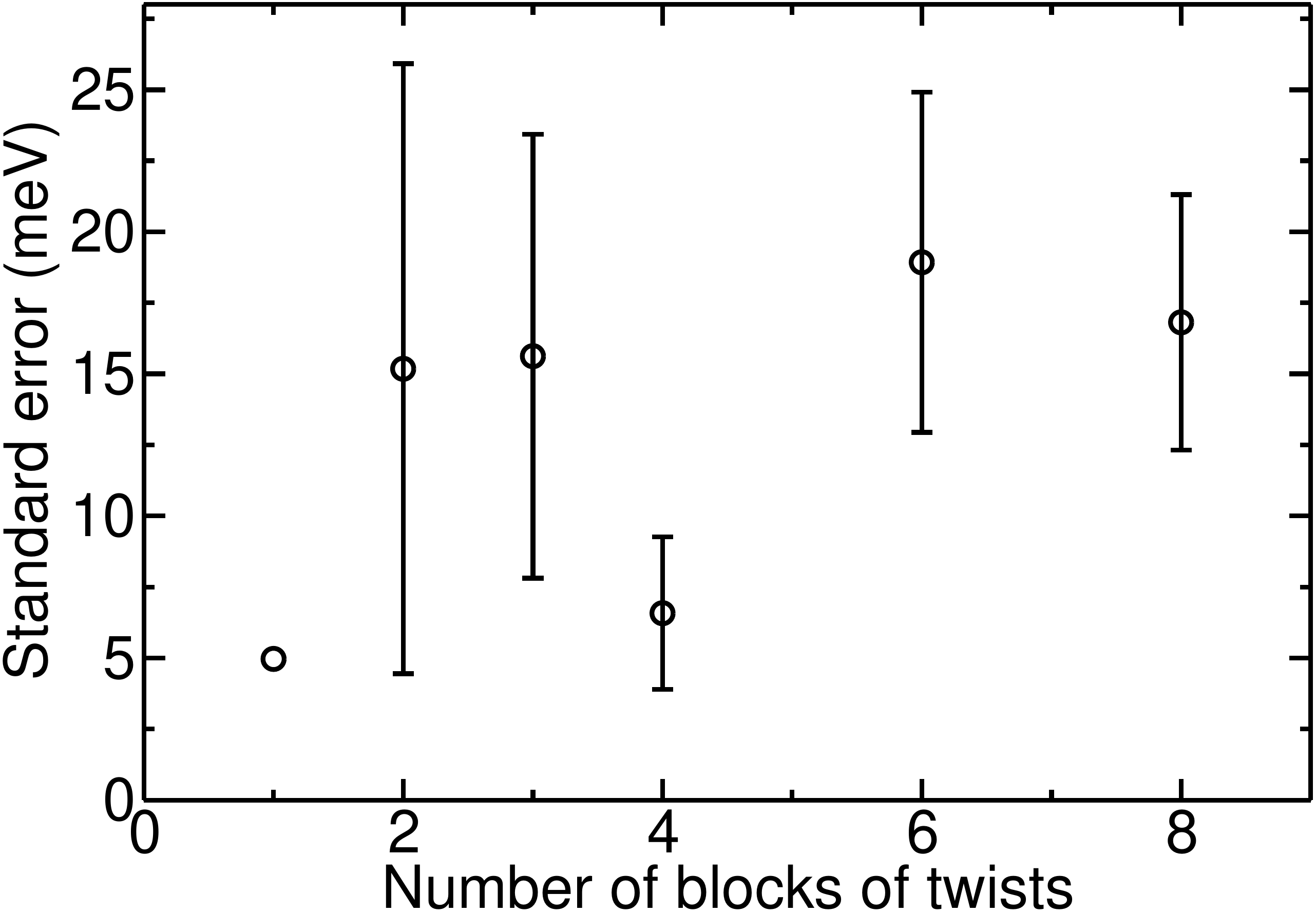}
\caption{TB standard error in the TA-DMC pure MV formation energy in a
  $3\times3$ supercell against the number of blocks into which the 24
  original twists are divided.
Within each block, Eq.\ \eqref{eq:tafit} is used to find the TA pure
formation energy.
The standard error in the single-block case is obtained by Gaussian
propagation of the Monte Carlo random errors, with no account being
taken of residual twist-sampling errors.}
\label{fig:twist_blocked_energy}
\end{figure}

Figure \ref{fig:twist_blocked_energy} shows the twist-blocked (TB)
standard error in the mean pure MV formation energy in a $3\times3$
supercell against the number of blocks into which the $24$ original
twists are divided.
Where there is just one block of twists the standard error is obtained
by Gaussian propagation of the Monte Carlo standard errors, with no
attempt to quantify the errors due to the random sampling of twists.
Due to the uncertainty in the estimated standard error,
Fig.\ \ref{fig:twist_blocked_energy} does not provide evidence that
there are significant ${\bf k}$-point sampling errors after fitting
Eq.\ \eqref{eq:tafit} to all $24$ twists.
Furthermore, there is no evidence to suggest that the random error
obtained by Gaussian propagation of the Monte Carlo errors in the fit
to all $24$ twists is unreliable.
The behavior of the TB standard error is similar for the other two
supercell sizes studied.

To our knowledge this is the first work to use twist-averaging to
evaluate a defect formation energy.
The approach is valid, since TA and non-TA finite-size energies agree
in the infinite-system-size limit.
Twist-averaging has the considerable advantage of greatly reducing
nonsystematic finite-size effects by turning a sum over supercell
reciprocal lattice vectors into an integral over ${\bf k}$, aiding
extrapolation to the thermodynamic limit.
For example, the standard deviations of the DMC pure MV formation
energies as functions of twist are $0.3$ eV, $0.2$ eV, and $0.1$ eV in
$3 \times 3$, $4\times 4$, and $5 \times 5$ supercells, respectively,
indicating the likely size of the quasirandom error in the pure defect
formation energy that would arise from the use of a non-TA
calculation.

\subsubsection{Long-range effects}

To deal with long-range finite-concentration and finite-size effects,
defect-formation energies have been calculated at various supercell
sizes $N$, where $N$ is the number of pristine primitive cells in the
supercell, and then extrapolated to infinite system size using an
appropriate scaling law.

In the case of the SiS, there is some charge transfer from the silicon
atom to the graphene sheet, giving the defect a dipole moment.
Defects in neighboring supercells lead to the inclusion of unwanted
electrostatic dipole-dipole interactions.
The screened interaction between charges in a 2D semiconductor is of
Rytova-Keldysh form \cite{Rytova_1965, Keldysh_1979}, which is
logarithmic at short range, before crossing over to a $1/r$
interaction at a lengthscale typically of order many tens of {\AA}\@.
The supercell sizes that we study here are comparable with this length
scale.
Rytova-Keldysh dipole-dipole interaction energies go as $r^{-3}$ at
long range and as $r^{-2}$ at short range; this leads to
finite-concentration errors that go as $O(N^{-1})$ in small
supercells, then as $O(N^{-3/2})$ in very large supercells.

The MV and SW defects are neutral and do not involve charge transfer
between atoms, so they have no dipole moment.
In principle there exists a quadrupole moment associated with these
defects, giving rise to weak electrostatic interactions between
periodic images falling off rapidly as $O(N^{-2}$--$N^{-5/2})$.

In addition to image interactions, there are elastic
finite-concentration effects due to the stress arising from the change
in the size and shape of the unit cells around a point defect in a
fixed supercell \cite{Castleton_2004}.
For example, if a point defect is associated with an area change
$\delta A$ and the supercell area is fixed at $NA$, and there is a
uniform, isotropic contraction of the lattice around the defect, the
resulting leading-order elastic energy is $(\lambda+\mu)\delta
A^2/(2AN)$, where $\lambda$ and $\mu$ are the Lam\'{e} coefficients of
graphene.
In general, assuming the defects result in isotropic stress, the
elastic finite-concentration effects in the energy go as $O(N^{-1})$.

\begin{figure}[!htpb]
\centering \includegraphics[scale=0.3,clip]{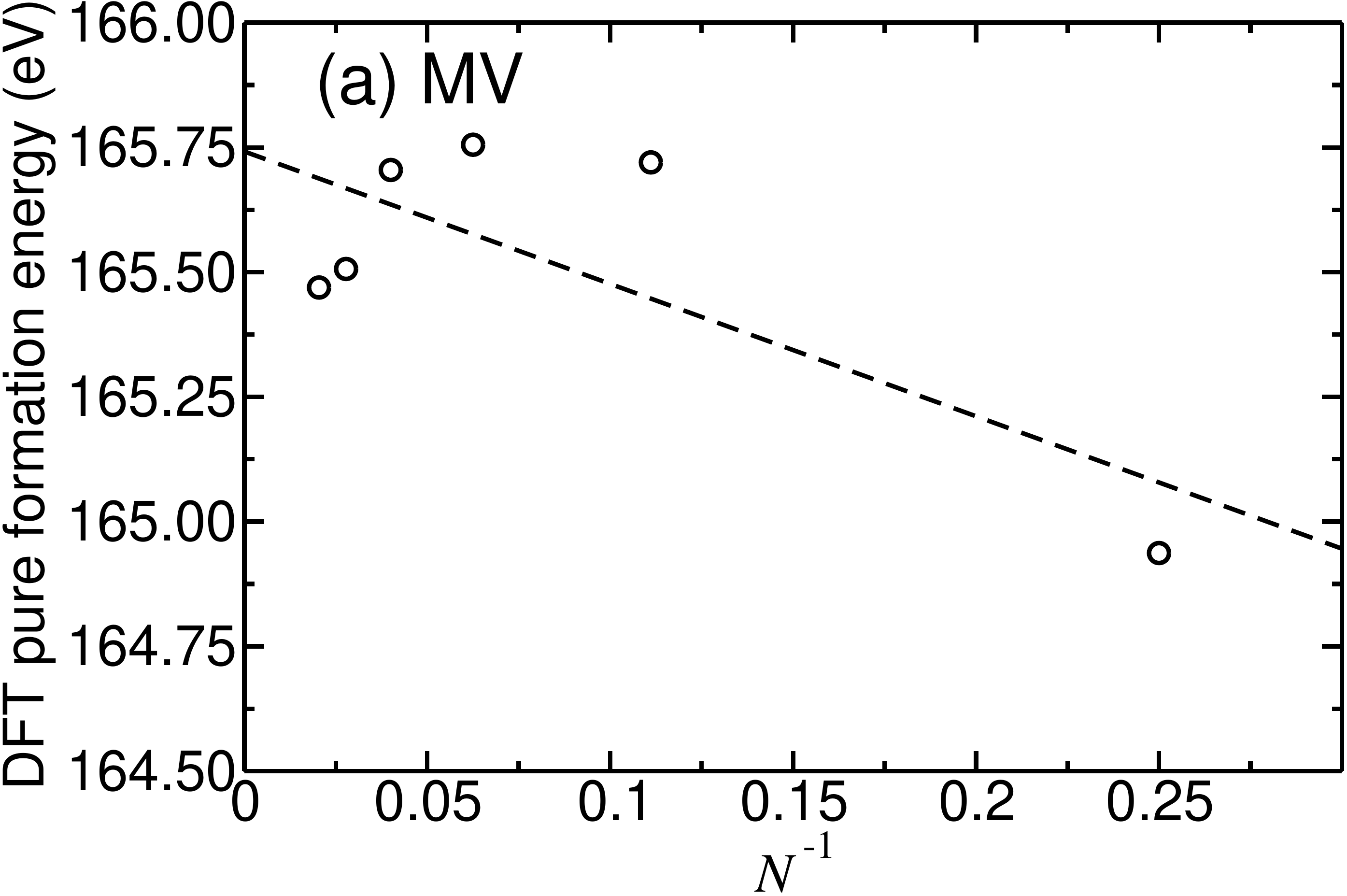}
\\ \includegraphics[scale=0.3,clip]{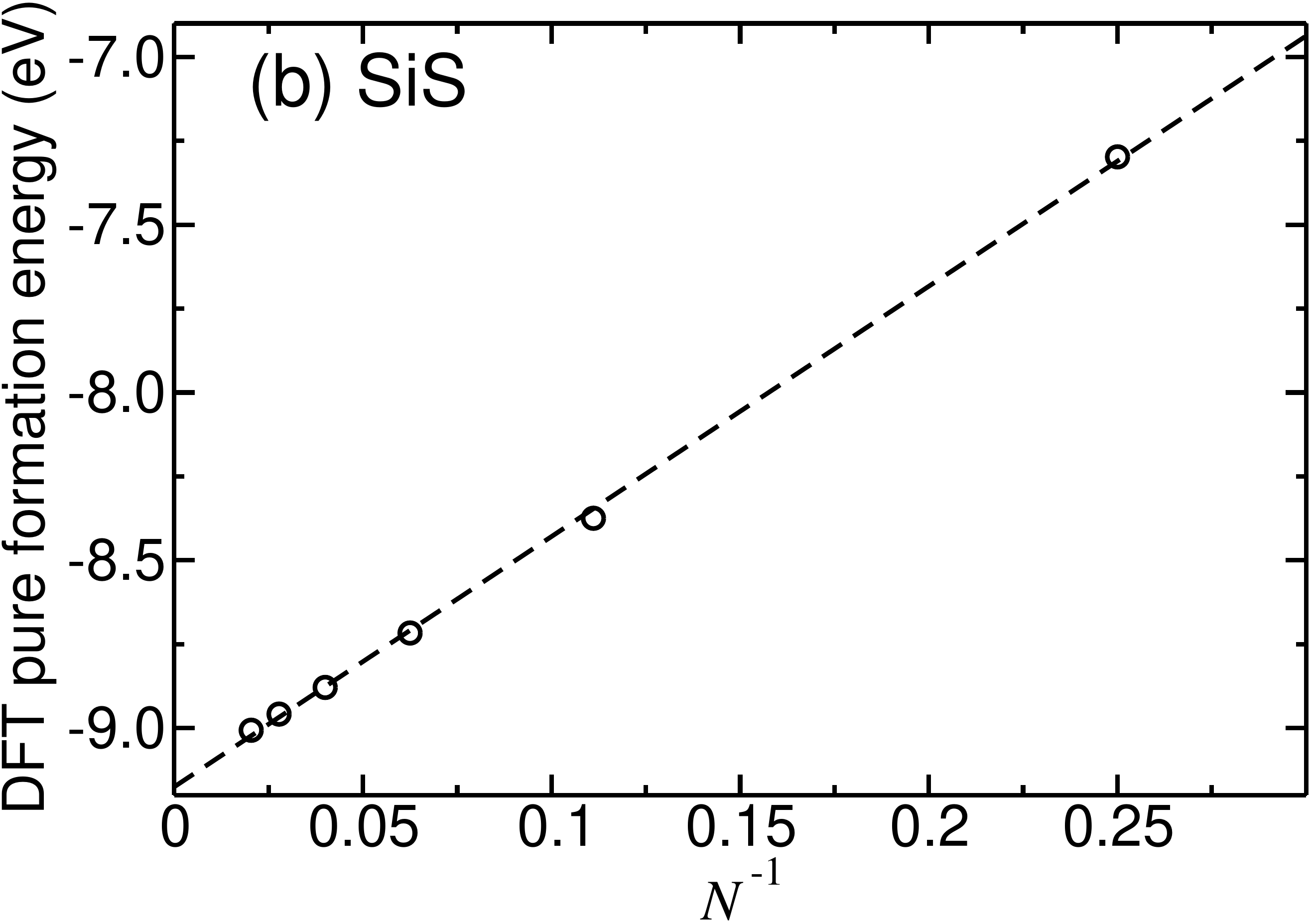}
\\ \includegraphics[scale=0.3,clip]{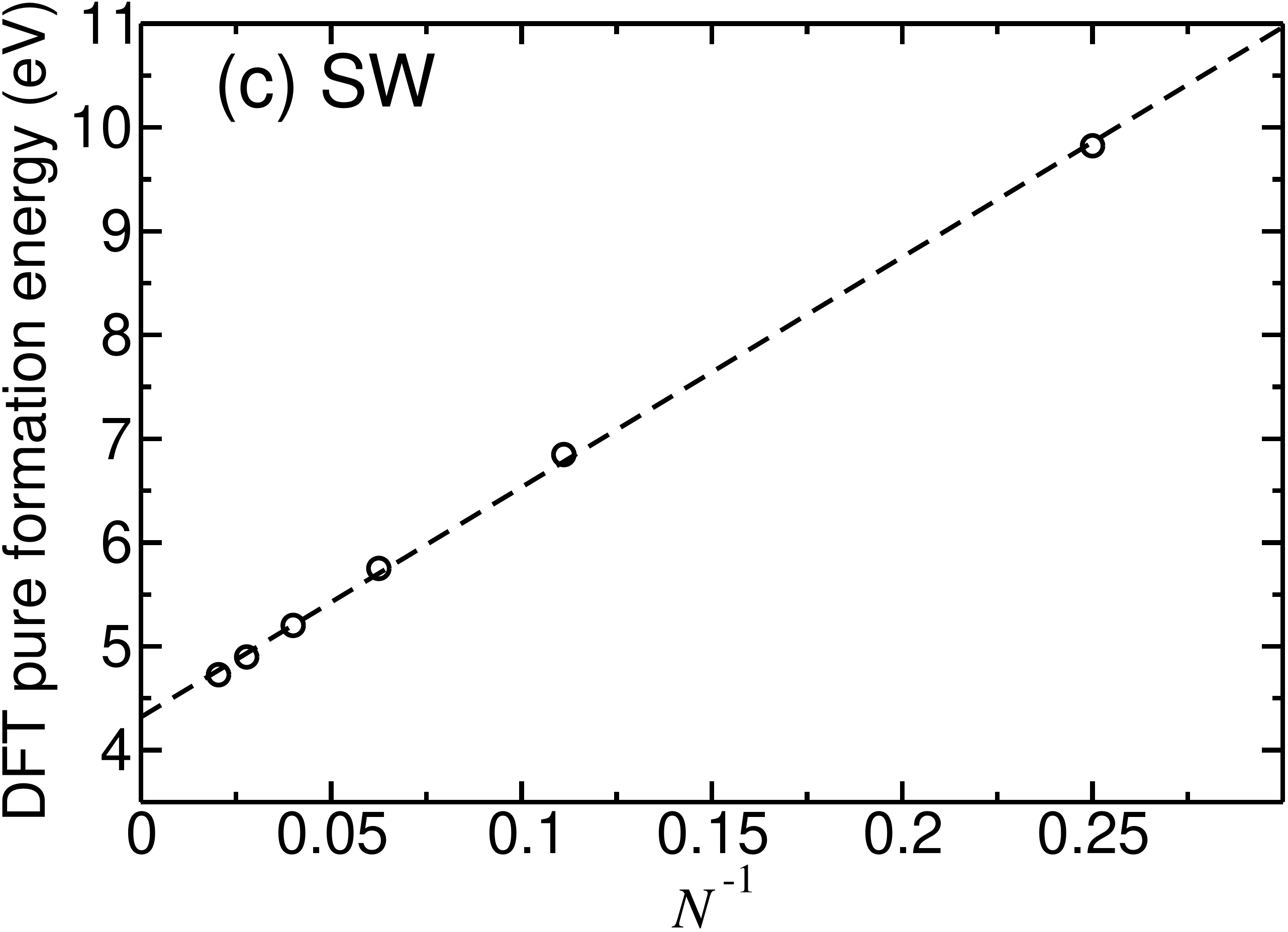}
\caption{DFT pure formation energies of (a) MV, (b) SiS, and (c) SW
  defects in graphene against the reciprocal of the supercell size $N$.
Fine ${\bf k}$-point grids were used in each supercell.
Ultrasoft pseudopotentials were used.
The dashed lines show fits of Eq.\ (\ref{eq:E_pf_fs_extrap}) to the
data.
\label{fig:dft_pure_formation}}
\end{figure}

In summary the scaling of the elastic finite-size error (and the
electrostatic finite-size error in the case of the SiS) suggest that
TA pure defect formation energies $\mathcal{E}^\text{pf}$ should be
extrapolated to the thermodynamic limit by fitting
\begin{equation} \mathcal{E}^\text{pf}(N)=\mathcal{E}^\text{pf}(\infty)+CN^{-1},
\label{eq:E_pf_fs_extrap}
\end{equation}
where $C$ is a fitting parameter.
Using DFT calculations, we confirm in
Fig.\ \ref{fig:dft_pure_formation} that $O(N^{-1})$ systematic
finite-concentration errors are dominant in MV, SW, and SiS defects in
graphene.

The pristine graphene, bulk silicon, and silicene energies per atom
were extrapolated to infinite system size by fitting the TA energies
per atom $\langle e_\text{P}(N)\rangle_\text{TA}$ to
\begin{eqnarray}
\langle e_\text{P}(N)\rangle_\text{TA} = e_\text{P}(\infty) +
c\:N^{-\gamma},
\end{eqnarray}
where $e_\text{P}(\infty)$ and $c$ are fitting parameters.
For graphene and silicene, $\gamma=5/4$ \cite{Drummond_2008}, while
for bulk silicon $\gamma=1$ \cite{Chiesa_2006}.

After separately dealing with the finite-size effects in the pure
defect formation energies and chemical potentials, graphene defect
formation energies were calculated using
Eqs.\ (\ref{eq:MV})--(\ref{eq:SW}).

% }}*

\subsection{Backflow \label{sec:backflow}}

DMC calculations in a $3\times3$ supercell at a single twist for the
MV show that the inclusion of backflow correlations with polynomial
electron-electron and electron-nucleus terms lowers the pure defect
formation energy by $41(30)$ meV, while the addition of the plane-wave
electron-electron term further lowers the pure defect formation energy
by $16(29)$ meV, giving a total lowering of $58(31)$ meV\@.
These differences are statistically insignificant, and are an order of
magnitude smaller than the error bars on the TA SJ-DMC pure defect
formation energies reported in Table
\ref{tab:defect_formation_energies}; fixed-node errors are therefore
well controlled.
A backflow function with electron-electron and electron-nucleus terms
lowers the energy per atom of graphene by $46(9)$ meV\@, and the
inclusion of the plane-wave electron-electron term does not have a
statistically significant effect.
Again, however, the effects of backflow are insignificant on the $0.1$
eV scale of the error bars on our SJ-DMC defect-formation energies.

% }}*

% *{{ Results and discussion

\section{Results and discussion}
\label{sec:results}

\subsection{Atomic structures}

\subsubsection{Pristine graphene, silicene, and bulk silicon}

We have used a carbon-carbon bond length of 1.42 {\AA} in all our
pristine graphene calculations \cite{Kelly1981, Dresselhaus1988}, and
we have used exactly the same supercell lattice parameters for our
pristine and defective graphene calculations.
For bulk silicon and silicene we used lattice parameters of $5.469$
and $3.866$ {\AA}, respectively, obtained using DFT with the
Perdew-Burke-Ernzerhof (PBE) exchange-correlation functional
\cite{Perdew_1996}.
The sublattice buckling of silicene ($0.458$ {\AA}) was also obtained
by relaxing within DFT-PBE \cite{drummond2012electrically}.

\subsubsection{MV}

It has previously been shown that a graphene MV undergoes a
Jahn-Teller distortion, with two neighbors of the missing atom moving
together to form a weak, reconstructed bond, and lowering the symmetry
from $D_{3h}$ point group \cite{ElBarbary_2003}.
One DFT work on MVs has found and used a $C_{2v}$ structure (a planar
structure with a single horizontal mirror plane and a single vertical
mirror plane) \cite{Wadey2016}; other DFT works have found that when
two neighbors of the missing atom form a reconstructed bond, the third
neighbor moves out of plane
\cite{ElBarbary_2003,Ronchi2017,Skowron2015}, resulting in a structure
of $C_s$ point group (a nonplanar structure with just a single
vertical mirror plane).

\begin{figure}[!htbp]
\centering
\subfloat{\includegraphics[scale=0.02,clip]{SC55mv_2.pdf}}
\qquad
\subfloat{\includegraphics[scale=0.02,clip]{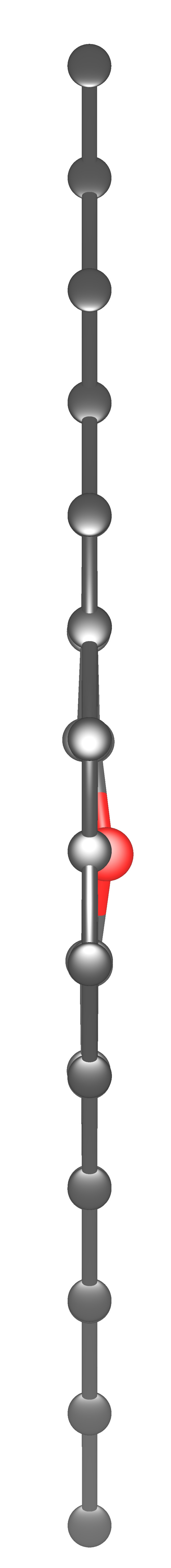}}
\caption{(a) Top-down and (b) in-plane views of the DFT-PBE-relaxed MV
  structure in a $5\times5$ supercell.
The undercoordinated carbon atom is shown in red.  }
\label{fig:MV_structure}
\end{figure}

As shown in Table \ref{table:DFT_MV_energies}, our non-spin-polarized
DFT calculations in the local density approximation (LDA) and DFT-PBE
calculations with and without many-body dispersion (MBD*) corrections
\cite{Tkatchenko_2012,Ambrosetti_2014} find the $C_s$ MV structure to
be favored (with the sole exception of DFT-LDA in a small, $3 \times
3$ supercell).
The energy differences between the different non-spin-polarized MV
structures oscillate significantly with supercell size, but are less
than $0.3$ eV\@.
This is just about large enough to be non-negligible on the scale of
our DMC error bars (see Sec.\ \ref{sec:formationresults}).
The difference between the $C_s$ DFT-PBE and DFT-PBE-MBD* structures
is small.
For example, in a $3\times 3$ supercell, the DFT-PBE energy is only
increased by $1.5$ meV when the DFT-PBE-MBD* structure is used instead
of the DFT-PBE structure.
We have used the non-spin-polarized $C_s$-symmetry structures obtained
by relaxing within DFT-PBE in our QMC calculations; the MV structure
in a $5 \times 5$ supercell is shown Fig.\ \ref{fig:MV_structure}.

\begin{table*}[!htpb]
\caption{DFT energies of the most stable $D_{3h}$, $C_{2v}$, and
  ${C_s}$ structures relative to the most stable non-spin-polarized
  ${C_s}$ structure.
Different exchange-correlation functionals and supercell sizes are
used.
Spin-polarized and non-spin-polarized results are labelled by ``(p)''
and ``(u)'', respectively.
Ultrasoft pseudopotentials and fine ${\bf k}$-point grids are used.
\label{table:DFT_MV_energies}}
\begin{ruledtabular}
\begin{tabular}{lccccc}

& & \multicolumn{4}{c}{Energy relative to non-spin-polarized $C_s$
    defect (meV)} \\

\raisebox{1ex}[0pt]{Functional} & \raisebox{1ex}[0pt]{Supercell} &
$E_{D_{3h}({\rm u})} - E_{C_s({\rm u})}$ & $E_{C_{2v}({\rm
    u})}-E_{C_s({\rm u})}$ & $E_{C_{2v}({\rm p})} - E_{C_s({\rm u})}$
& $E_{C_s({\rm p})}-E_{C_s({\rm u})}$ \\

\hline

LDA & $3\times 3$ & $0$ & $0$ & $0$ & $0$ \\

LDA & $5\times 5$ & $94.6$ & $92.4$ & $71.5$ & $0$ \\

LDA & $7\times 7$ & $120.2$ & $115.4$ & $88.7$ & $0$ \\

PBE & $3\times 3$ & $21.5$ & $21.5$ & $-36.4$ & $-36.5$ \\

PBE & $5\times 5$ & $214.8$ & $211.6$ & $-96.8$ & $-96.8$ \\

PBE & $7\times 7$ & $274.2$ & $268.6$ & $85.6$ & $-113.8$ \\

PBE-MBD* & $3\times 3$ & $19.6$ & $19.6$ & $15.4$ & $19.8$ \\

PBE-MBD* & $5\times 5$ & $195.6$ & $193.7$ & $-105.7$ & $0$ \\

PBE-MBD* & $7\times 7$ & $245.9$ & $250.8$ & $-128.0$ & $-122.3$ \\

\end{tabular}
\end{ruledtabular}
\end{table*}

Previous DFT calculations have found that the MV has a magnetic moment
of $1.04\mu_\text{B}$--$2\mu_\text{B}$, where $\mu_\text{B}$ is the
Bohr magneton \cite{Ma_2004,Paz2013,Valencia_2017}.
We examine the effect of performing spin-polarized DFT calculations in
Table \ref{table:DFT_MV_energies}.
Within DFT-LDA, the MV is unambiguously nonmagnetic.
Within DFT-PBE and DFT-PBE-MBD*, spin-polarized structures of $C_{2v}$
and $C_s$ symmetry are found to be stable.
Convergence to the lowest-energy atomic structure is challenging in
spin-polarized DFT calculations, where we often find structures with
energies that are either greater than the non-spin-polarized energy
for the same point group or greater than the energy of a
higher-symmetry structure, demonstrating that we have not obtained the
global minimum of the energy.
To try to address this problem we have performed repeated
spin-polarized calculations with different initial plane-wave
coefficients and different initial geometries.
Our DFT-PBE calculations in a $7 \times 7$ supercell suggest that the
spin-polarized $C_s$ MV structure is more stable than the
non-spin-polarized $C_s$ structure by about $0.1$ eV, in agreement
with previous DFT calculations \cite{Ma_2004}, with magnetic moment
$1.4\mu_\text{B}$.
However, the energy differences between magnetic and nonmagnetic MV
structures are less than or comparable to the error bars on our DMC
defect-formation energies reported in
Sec.\ \ref{sec:formationresults}.
Furthermore, there is no sign of convergence with respect to supercell
size of the difference between the DFT energies of the magnetic and
nonmagnetic structures.
In a $3\times3$ supercell, the DMC pure formation energy obtained
using spin-polarized orbitals is within error bars of that obtained
with non-spin-polarized orbitals.
For consistency, we have used nonmagnetic MV structures and
non-spin-polarized orbitals in our QMC calculations.

\subsubsection{SiS}

\begin{figure}[!htbp]
\centering
\subfloat{\includegraphics[scale=0.041,clip]{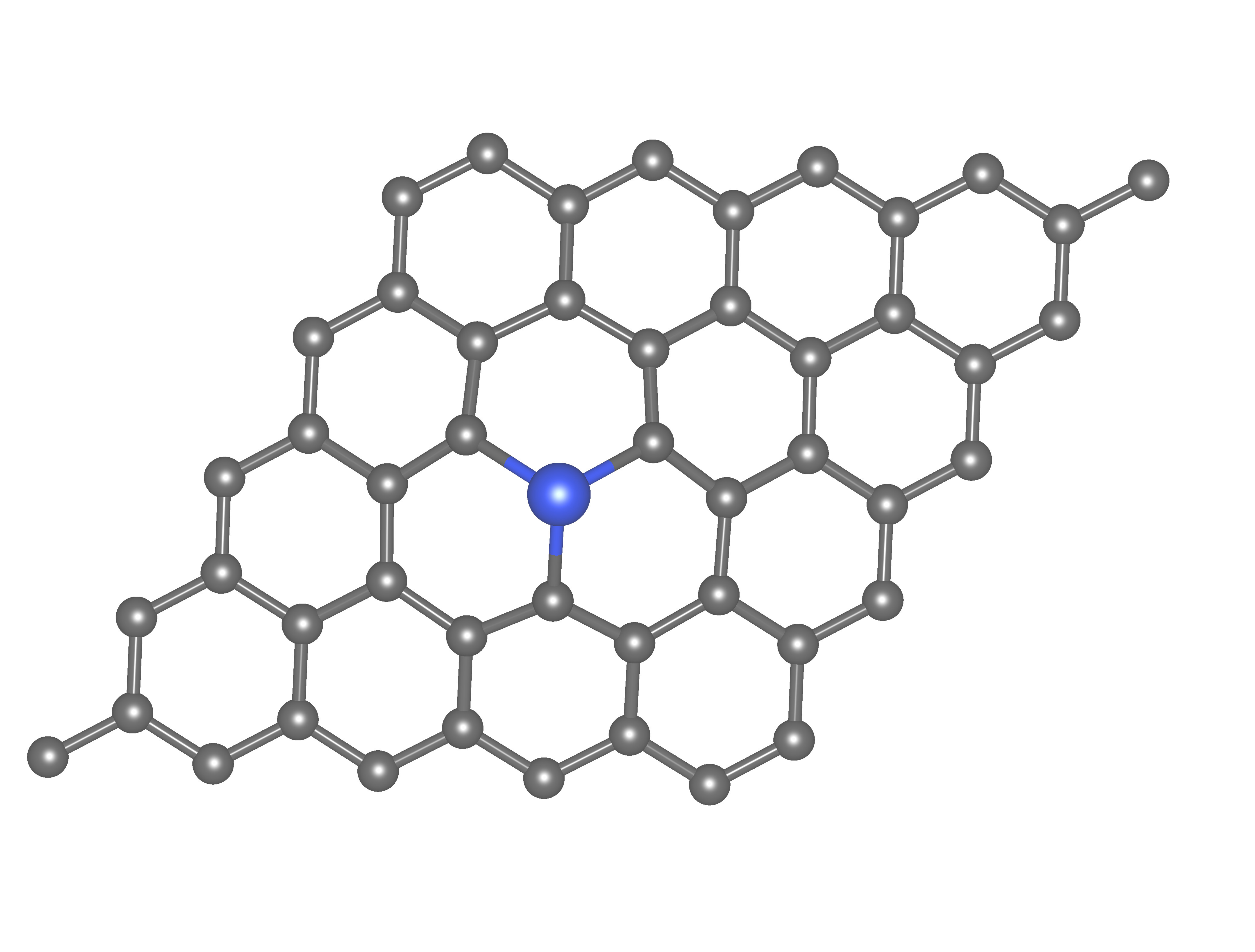}}
\qquad
\subfloat{\includegraphics[scale=0.032,clip]{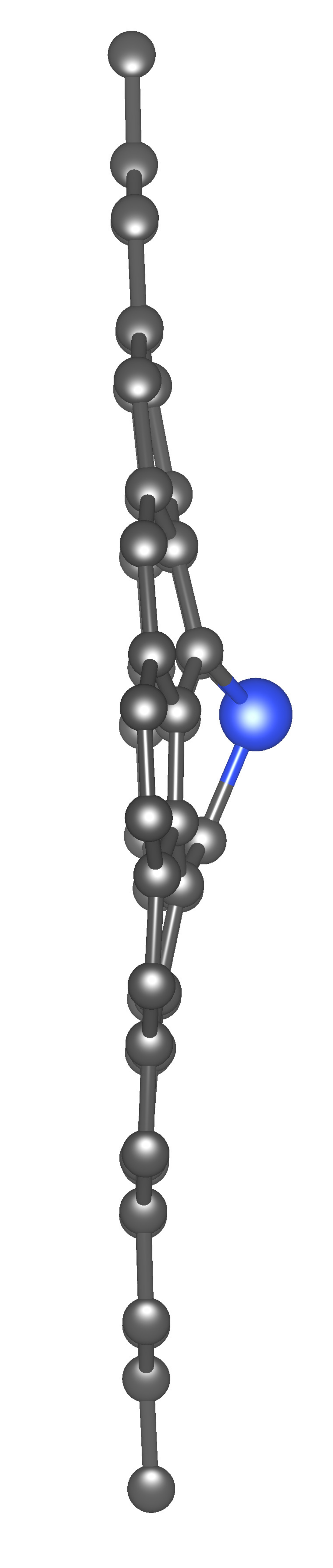}}
\caption{(a) Top-down and (b) in-plane views of the DFT-PBE-relaxed
  SiS structure in a $5\times5$ supercell.
The silicon atom is shown in blue.}
\label{fig:SiS_structure}
\end{figure}

Replacing a single carbon atom by a silicon atom results in a defect
of $C_s$ point group, rather than $D_{3h}$, due to a Jahn-Teller
distortion \cite{banhart2010structural}.
The DFT-PBE-relaxed geometry shown in Fig.\ \ref{fig:SiS_structure}
has the silicon atom bonded with three carbon atoms and lying above
the graphene plane due to partial $sp^3$ hybridization.

\subsubsection{SW defect}

\begin{figure}[!htbp]
\centering
\subfloat[\label{subfig:sw1}]{\includegraphics[scale=0.049,clip]{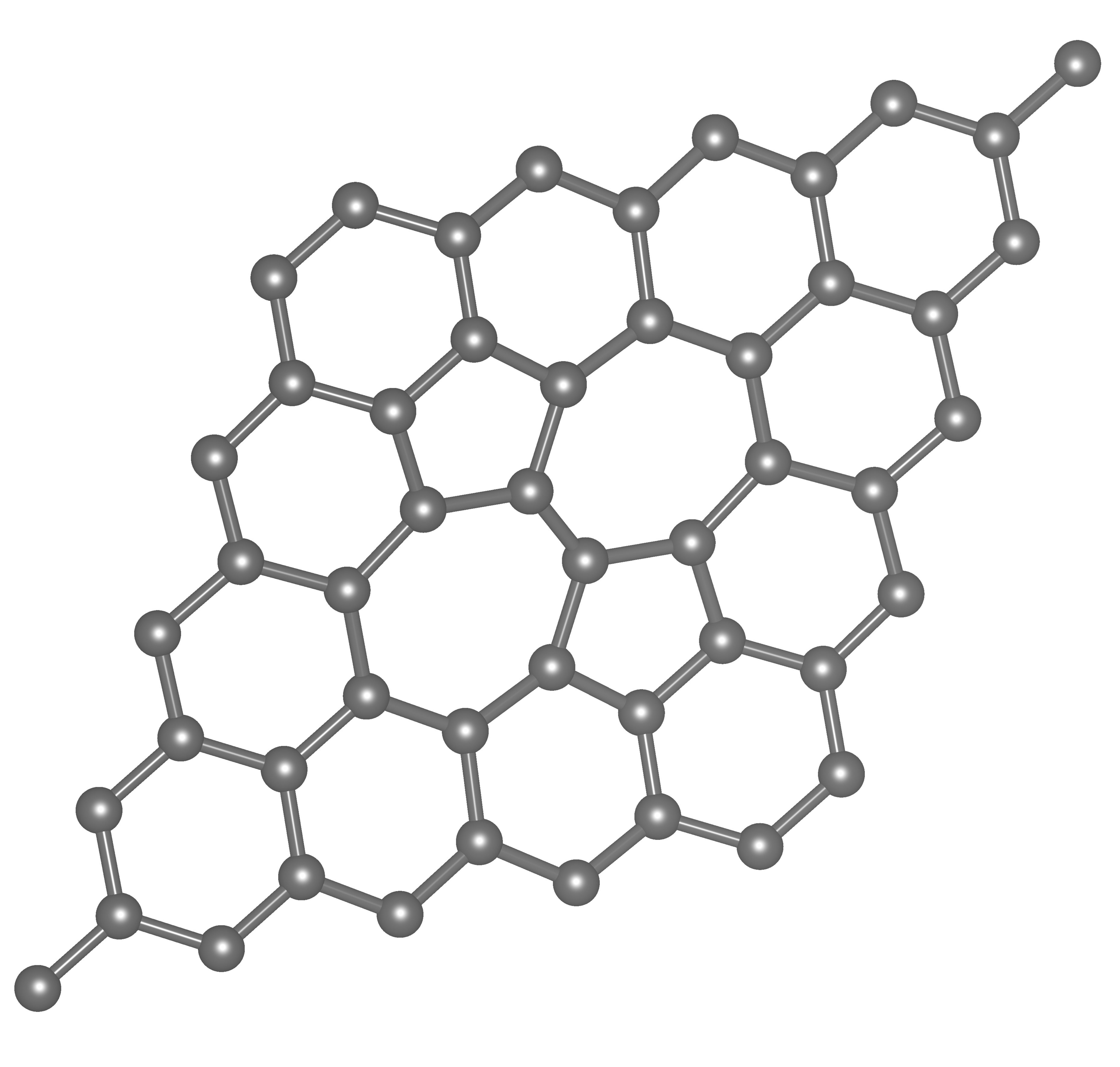}}
\subfloat[\label{subfig:sw2}]{\includegraphics[scale=0.031,clip,angle=90]{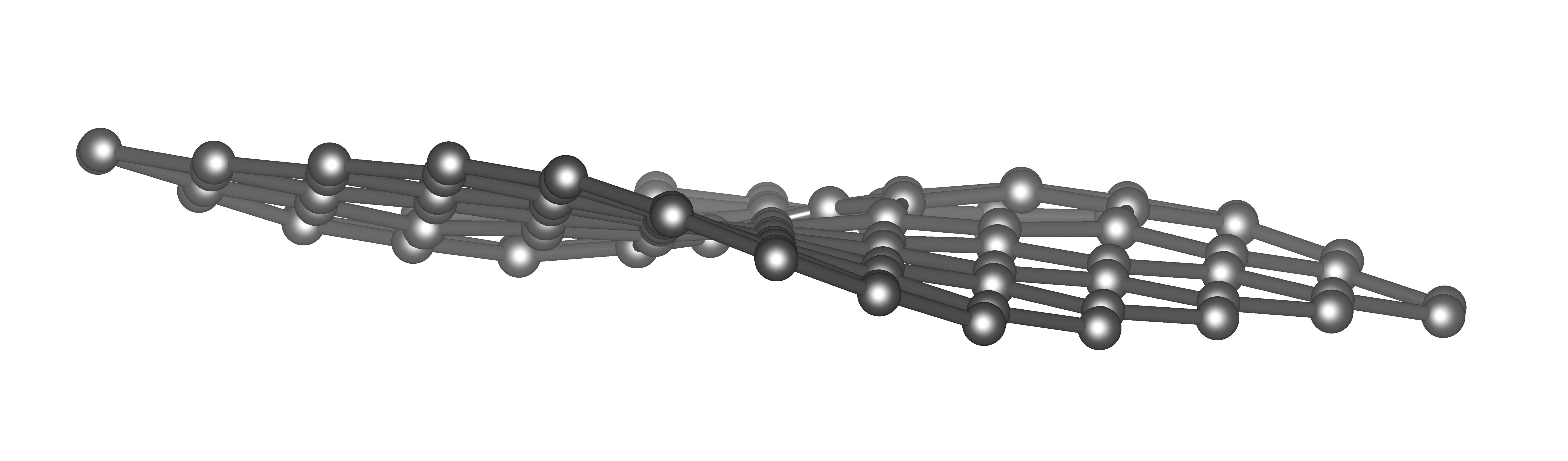}}
\caption{(a) Top-down and (b) in-plane views of the DFT-PBE-relaxed
  ``sine-like'' SW defect structure in a $5\times5$ supercell.}
\end{figure}

%structure
In graphene, a SW defect is formed by an in-plane rotation of a single
carbon-carbon bond through $90^{\circ}$ about its midpoint.
This transforms four hexagonal unit cells into two pentagons and two
heptagons, as shown in Fig.\ \subref*{subfig:sw1}, with the same
number of carbon atoms as pristine graphene and without any dangling
bonds.
The SW rotation compresses or stretches many bonds, resulting in a
wave of significant vertical displacement of carbon atoms around the
defect, as shown in Fig.\ \subref*{subfig:sw2}.
The relaxed lattice will adopt either a ``sine-like'' buckled
structure, in which the two rotated carbon atoms are slightly
displaced in opposite out-of-plane directions, or a ``cosine-like''
buckled structure, in which the two rotated carbon atoms are slightly
displaced in the same out-of-plane direction.
The ``sine-like'' structure is the lower-energy configuration
\cite{guedj2018atomistic,ma2009stone}, and is the structure studied in
this work.

\subsection{Defect formation energies}
\label{sec:formationresults}

\begin{figure}[!htpb]
\centering
\includegraphics[scale=0.3,clip]{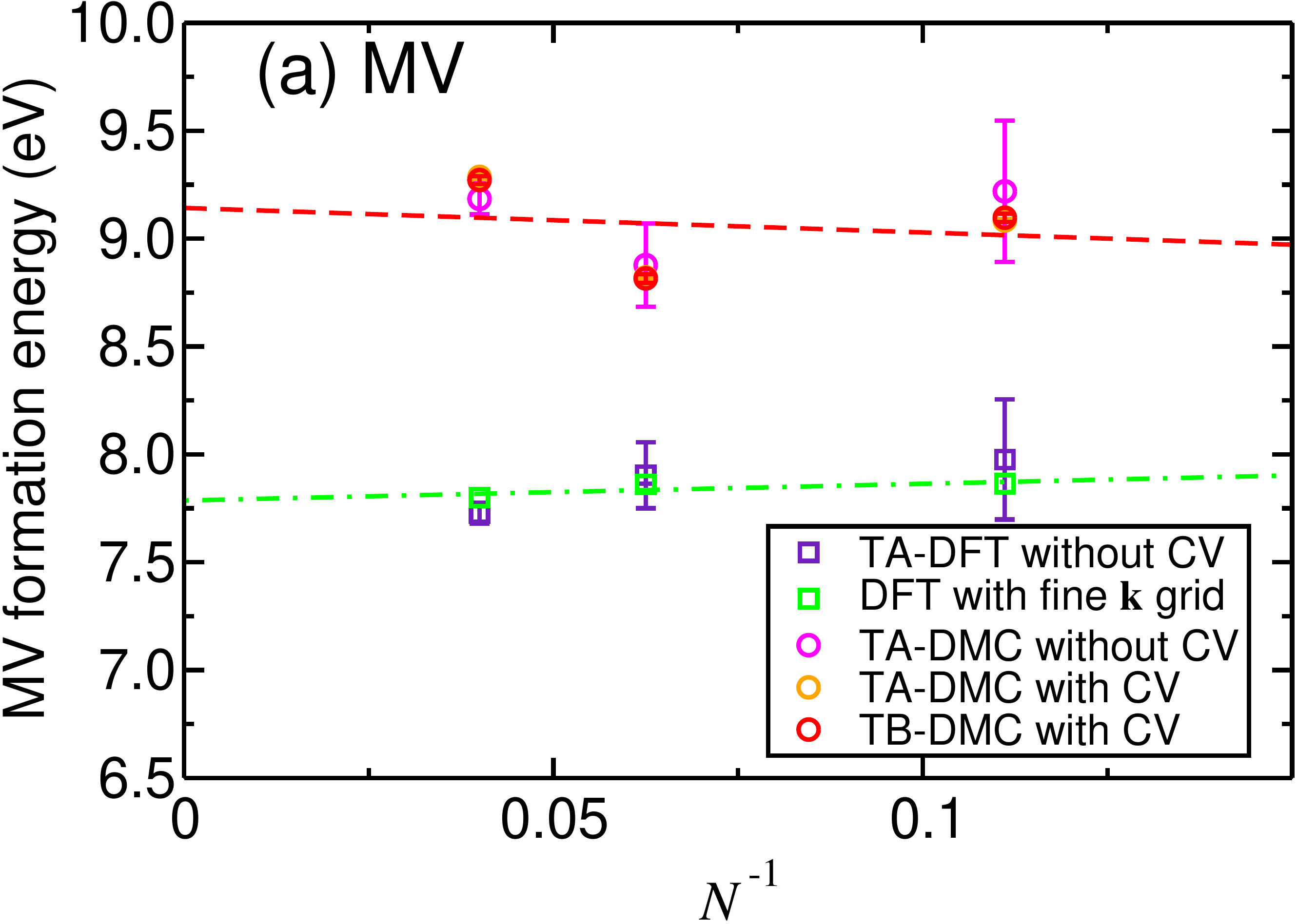}
\\ \includegraphics[scale=0.3,clip]{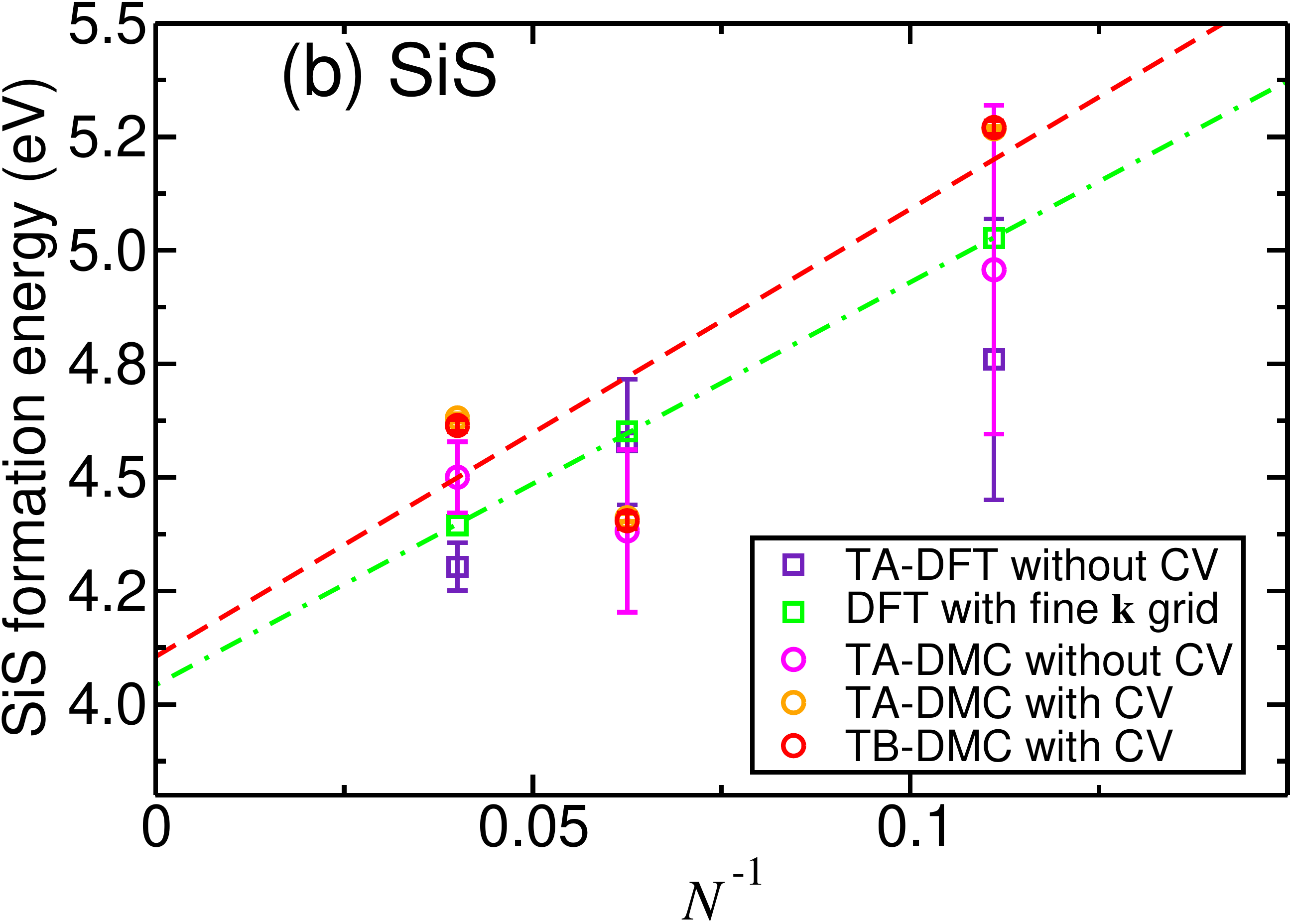}
\\ \includegraphics[scale=0.3,clip]{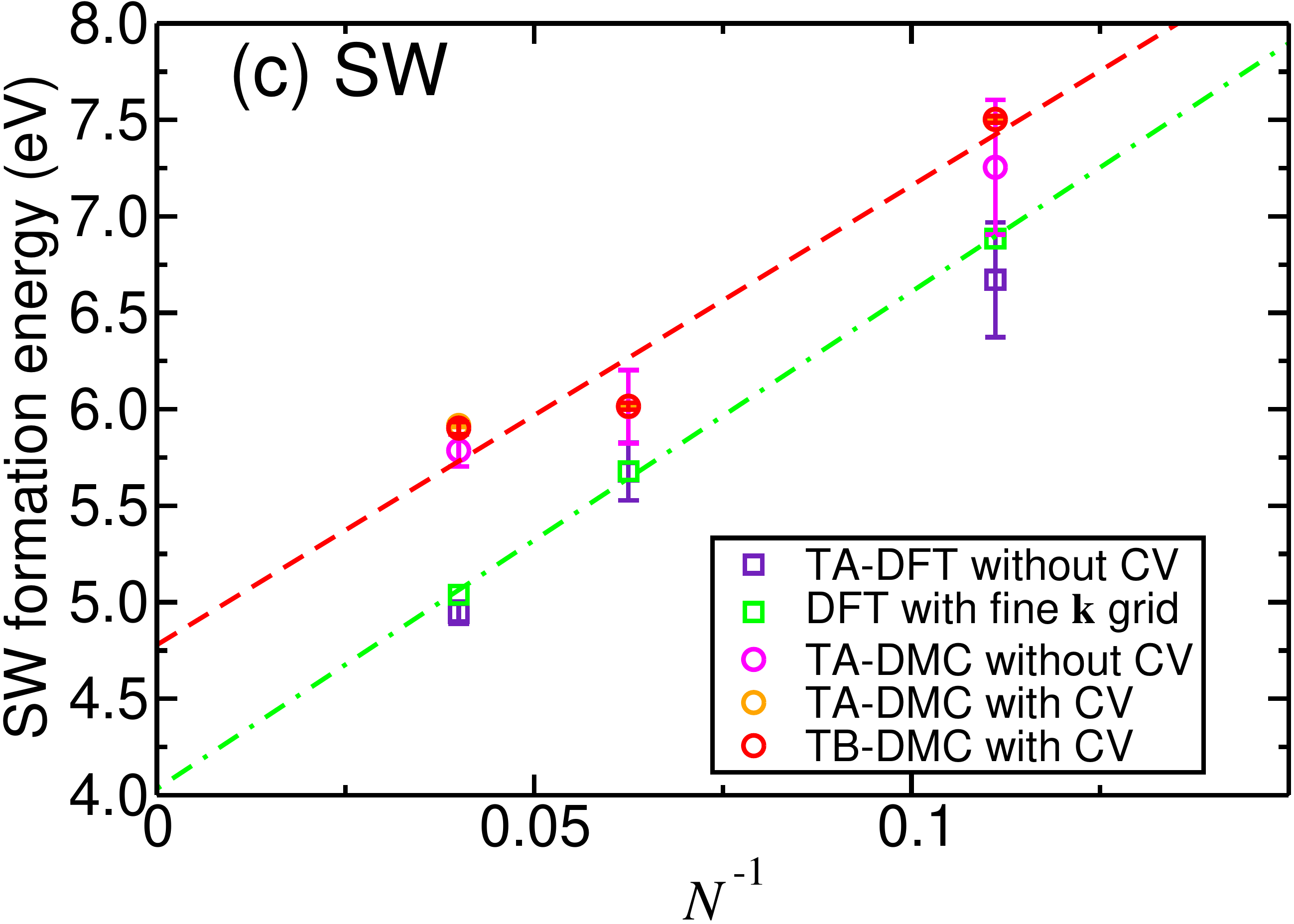}
\caption{DFT and DMC formation energies against reciprocal of
  supercell size $N$, using different methods for dealing with
  momentum quantization errors, for (a) MV, (b) SiS, and (c) SW
  defects.
  The red dashed lines show an unweighted least-squares fit of
  Eq.\ (\ref{eq:E_pf_fs_extrap}) to the TA-DMC data.
Both the DFT and DMC calculations used Dirac-Fock pseudopotentials.
The carbon and silicon chemical potentials were taken to be the energy
per atom of monolayer graphene and bulk silicon, extrapolated to
infinite system size; hence the $N$-dependence shown in this figure
only arises from the finite-concentration and finite-size effects in
the pure formation energy.}
\label{fig:FormationEnergyComp}
\end{figure}

\begin{table*}[!htpb] \caption{\label{tab:defect_formation_energies}
Theoretical static-nucleus formation energies for various point
defects in monolayer graphene.
The carbon and silicon chemical potentials are the energies per atom
of graphene and bulk silicon, respectively.
Results without citations were obtained in the present work.
``NTBM'' refers to a nonorthogonal tight-binding model.
``DFT-B3LYP-D*'' refers to DFT with a self-interaction-corrected
hybrid B3LYP functional \cite{Lee_1988,Becke_1993}.
To compare with experimental results, the vibrational free energies
reported in Table \ref{tab:vib_en} should be added to the
static-nucleus data reported in this table.}
\begin{ruledtabular} \begin{tabular}{lcccc}

\multirow{2}{*}{Method}& \multicolumn{3}{c}{Defect formation energy
  (eV)} \\

& MV & SiS & SW \\

\hline

DFT-PBE & $7.64$ \cite{Wadey2016}, $7.65$ \cite{ervasti2015silicon},
$7.97$& $3.77$ \cite{ervasti2015silicon}, $6.85$
\cite{Denis2010}\footnote{Reference \onlinecite{Denis2010} uses the
  ground-state energies of isolated atoms as chemical potentials; for
  comparison with the other defect formation energies reported in this
  table, the atomization energies of graphene and bulk silicon should
  be, respectively, added to and subtracted from the formation energy
  of Ref.\ \onlinecite{Denis2010}.}  $3.59$ & $4.71$
\cite{ma2009stone}, $4.32$ \\

DFT-LDA  & $8.02$ \cite{Wadey2016}, $7.40$
\cite{zobelli2012comparative}, $8.25$ & & $4.66$
\cite{shirodkar2012electronic},\footnote{This work extrapolates DFT
  energies at different system sizes in the same fashion we do here.
  All other cited DFT works are performed at finite supercell size.} $4.86$
\cite{zobelli2012comparative}, $5.42$ \cite{ma2009stone}\\

DFT-B3LYP-D* & $8.05$ \cite{Ronchi2017}& & \\

NTBM & & &$4.60$ \cite{podlivaev2015out} \\

DMC & & & $5.82(3)$ \cite{ma2009stone} \\

DMC-corrected DFT & $9.0(1)$ & $4.4(1)$ & $4.9(1)$

\end{tabular}
\end{ruledtabular}
\end{table*}

Figure \ref{fig:FormationEnergyComp} shows DMC and DFT defect
formation energies against system size.
Twist averaging is either performed directly (without a CV), or by
using the DFT results as a CV [i.e., fitting Eq.\ (\ref{eq:tafit})].
Also shown are DFT results obtained with a fine ${\bf k}$-point mesh.
The error bars on the ``TA-DMC with CV'' data were obtained using
Gaussian propagation of errors through the fit of Eq.\
(\ref{eq:tafit}) to the DMC results at all 24 twists.
The ``TB-DMC'' data were obtained using twist-blocking, in which six
blocks of four twists were used to obtain a standard error estimate
that includes both Monte Carlo random errors and
finite-twist-sampling random errors.
Figure \ref{fig:FormationEnergyComp} demonstrates that the use of a
CV significantly reduces the random errors in the DMC energy data,
and that subsequent twist-blocking to account for the remaining
twist-sampling errors does not affect the random error estimate
significantly (as we have also shown in Fig.\
\ref{fig:twist_blocked_energy}).
In theory, the most accurate way to obtain the TA energy is to fit
Eq.\ (\ref{eq:tafit}) to formation energies in a single block of all
the twists, and then to use TB to obtain the error bars, but the
difference between the TA and TB mean energies is negligible in
practice.
However, the TB errors are not large enough to quantify the
quasirandom finite-size errors in the formation energies at different
supercell sizes; this finite-size noise must therefore arise from
effects such as the enforced supercell commensurability of
Ruderman-Kittel oscillations in the density and pair density rather
than momentum quantization.
Quasirandom finite-size effects are larger in the DMC formation
energies than in the DFT results, presumably because of the explicit
treatment of correlation in QMC methods.
The obvious (but expensive) way to reduce this would be to perform
DMC calculations in a larger range of supercell sizes and possibly
shapes.

At each system size we evaluate a correction to the DFT formation
energy as a difference between the TA-DMC result and the DFT result
with a fine ${\bf k}$-point grid.
The DMC corrections to the defect formation energies in different
supercells (including the chemical potentials extrapolated to the
thermodynamic limit) are given in Table \ref{tab:DMCcorrections}.
In general, the difference between the DFT and DMC formation energies
is expected to be dominated by short-range effects, with systematic
finite-concentration errors (due to electrostatic and elastic effects)
being similar in DFT and DMC; this is confirmed by the similar
gradients of the fitted lines in
Fig.\ \ref{fig:FormationEnergyComp}.
However, the difference between DFT and DMC shows quasirandom
fluctuations as a function of system size.
This suggests that the best scheme for using DMC to evaluate defect
formation energies is to average the difference between TA-DMC and
fine-${\bf k}$-point DFT formation energies obtained in multiple
supercells, and then to apply the resulting correction to DFT results
extrapolated to the dilute limit of infinite supercell size.
Averaging over multiple supercells is clearly necessary, because the
difference between DMC and DFT results obtained in different cell
sizes in Table \ref{tab:DMCcorrections} fluctuates randomly by an
amount that is significantly larger than the error bars on the
individual differences.
DFT-PBE significantly underestimates the formation energy for all
three defects.
The larger DMC correction for the MV formation energy compared to the
SiS and SW defects reported in Table \ref{tab:DMCcorrections} suggests
that DFT performs relatively poorly when evaluating energy differences
between structures with very different chemical bonding.

\begin{table}[!htpb] \caption{\label{tab:DMCcorrections} Difference
between DMC and DFT-PBE static-nucleus formation energies for various
defects, evaluated as the difference between defect-formation
energies calculated with TA-DMC [using Eq.\ (\ref{eq:tafit})] and
DFT-PBE using a fine ${\bf k}$-point grid.
The DFT calculations use ultra-soft pseudopotentials rather than the
Trail-Needs Dirac-Fock pseudopotentials used by the QMC calculations.
}
\begin{ruledtabular}
\begin{tabular}{lccc}

\multirow{2}{*}{Supercell} & \multicolumn{3}{c}{\makebox[0pt]{DMC correction to
  formation energy (eV)}} \\

& MV & SiS & SW \\

\hline

$3\times3$ & $1.09(3)$ & $1.07(9)$ & $0.65(2)$ \\

$4\times4$ & $0.78(3)$ & $0.54(9)$ & $0.27(2)$ \\

$5\times5$ & $1.28(2)$ & $0.91(9)$ & $0.70(3)$ \\

\hline

Mean & $\!\!\!1.1(1)$ & $\!\!\!0.8(1)$ & $\!\!\!0.5(1)$ \\

\end{tabular}
\end{ruledtabular}
\end{table}

Our final DFT and DMC defect formation energies are shown in Table
\ref{tab:defect_formation_energies}, along with DFT results from the
literature.
The DMC literature result for the SW formation energy
\cite{ma2009stone}, which has a comparatively tiny standard error, was
only calculated in a $5\times5$ supercell with no attempt to control
finite-size effects; our standard error is larger because it accounts
for quasirandom finite-size effects.

The DFT-PBE differences in the zero-point vibrational energies and
Helmholtz free energies at $298$ K between defective and pristine
graphene for the MV, SiS, and SW defects are shown in Table
\ref{tab:vib_en}.
These vibrational free energy contributions should be added to the
static-nucleus defect formation energies in Table
\ref{tab:defect_formation_energies}.
The vibrationally corrected DMC defect formation energies are
$8.3(1)$, $3.6(1)$, and $4.4(1)$ at $298$ K for MV, SiS, and SW
defects, respectively.
The vibrationally corrected DMC MV formation energy may be compared
with an experimentally determined MV formation energy, $7.0(5)$ eV
\cite{Thrower1978}.
While the difference between this experimental result and our result is
statistically significant, it should be noted that the uncertainty in
the experimental result is fairly large.
Furthermore, there are some systematic errors that affect our DMC
result, such as the use of DFT-relaxed geometries and estimating the
Helmholtz free energy using the harmonic approximation.

\begin{table}[!htpb]
\caption{DFT-PBE vibrational contributions to the Helmholtz free
  energies of formation of various point defects in monolayer
  graphene. The contributions are extrapolated to the dilute limit.}
\label{tab:vib_en} \begin{ruledtabular}
\begin{tabular}{lccc}

\multirow{2}{*}{Temperature (K)} &
\multicolumn{3}{c}{\makebox[0pt]{Vib.\ contrib.\ to form.\ energy (eV)}} \\

& MV & SiS & SW \\

\hline

0 & $-0.74$ & $-0.44$ & $-0.49$ \\

298 & $-0.68$ & $-0.41$ & $-0.47$ \\

\end{tabular}
\end{ruledtabular}
\end{table}

\subsection{Atomization energies}

\begin{figure}[!htbp]
\centering
\includegraphics[scale=0.3,clip]{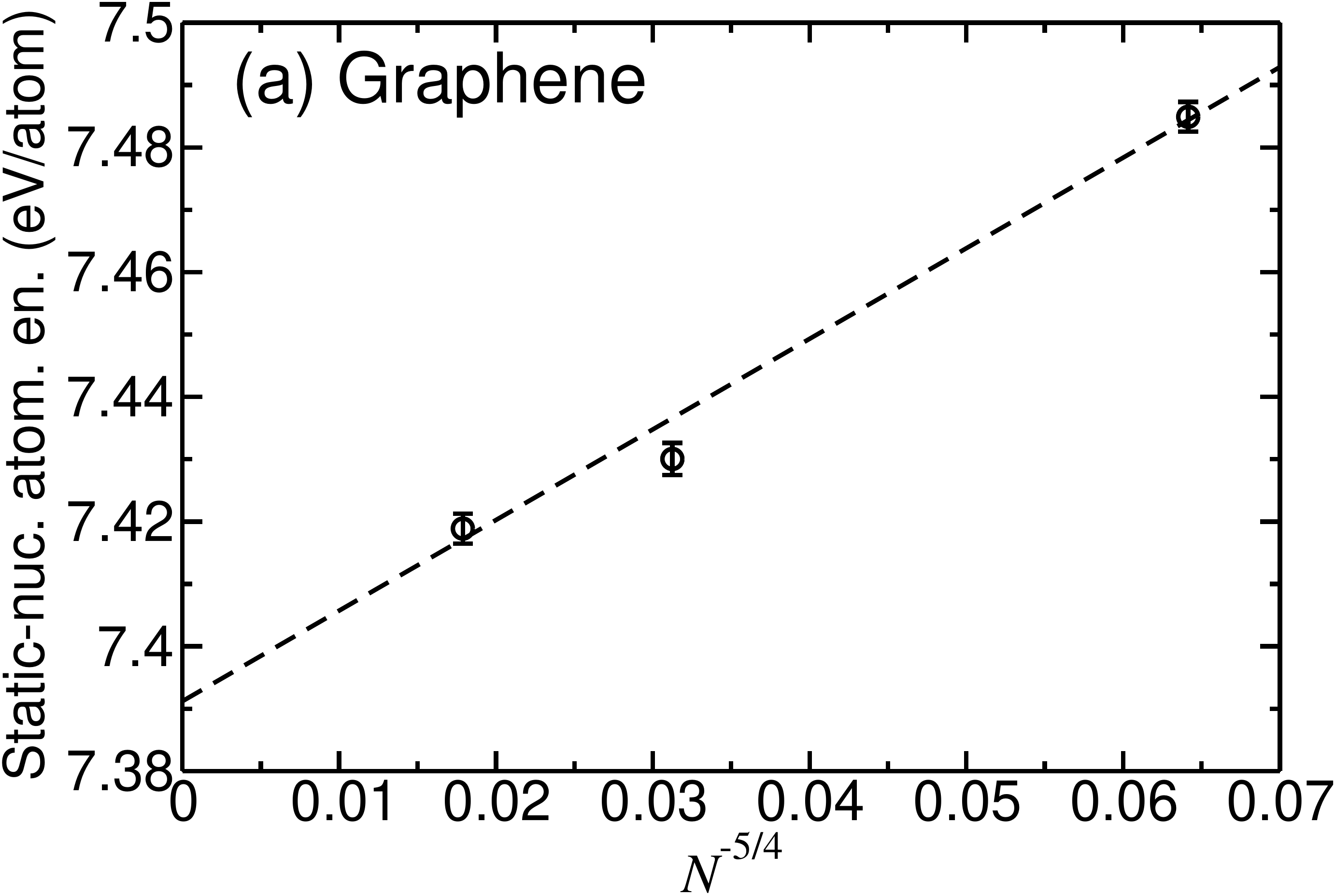}
\includegraphics[scale=0.3,clip]{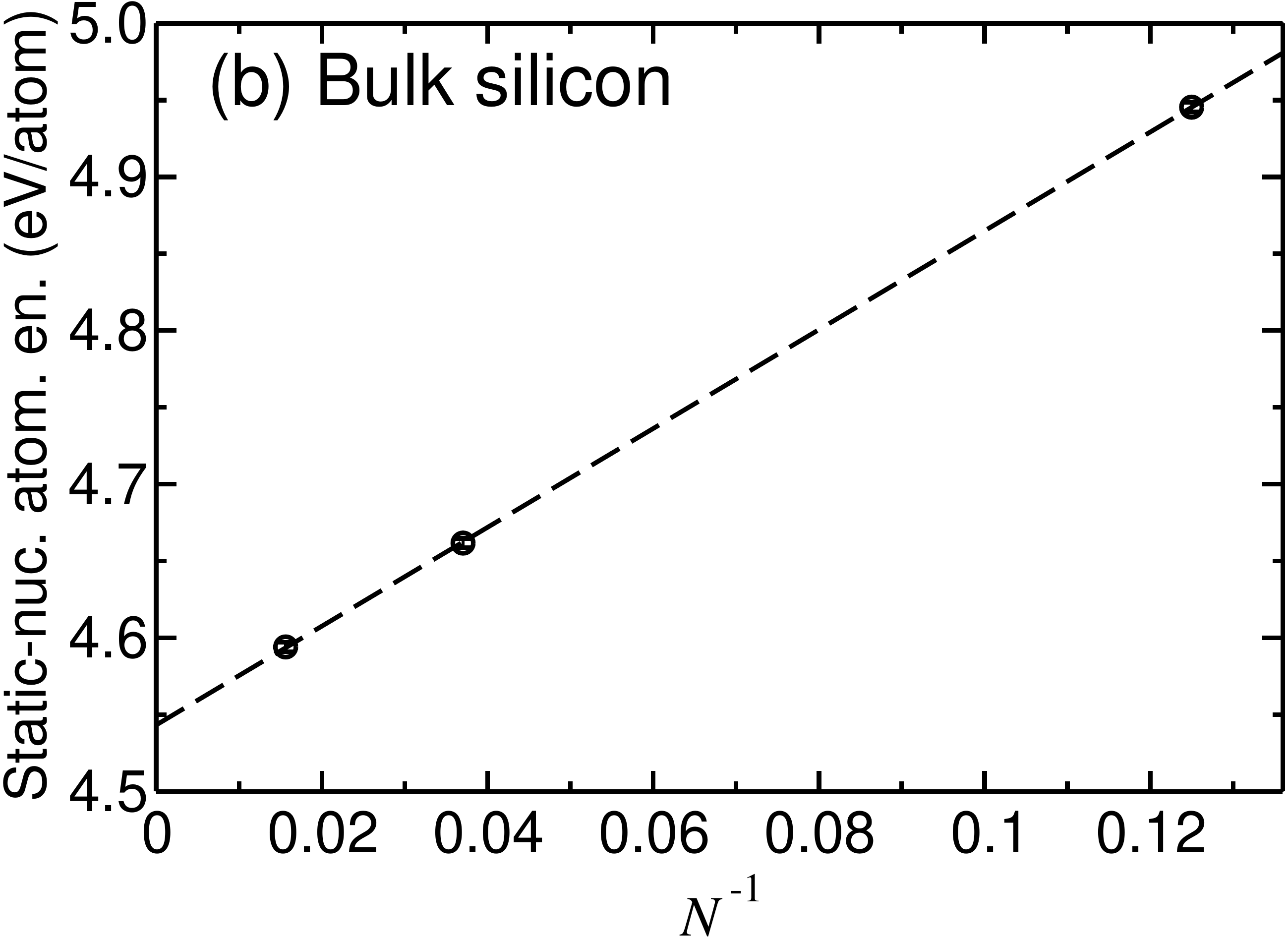}
\includegraphics[scale=0.3,clip]{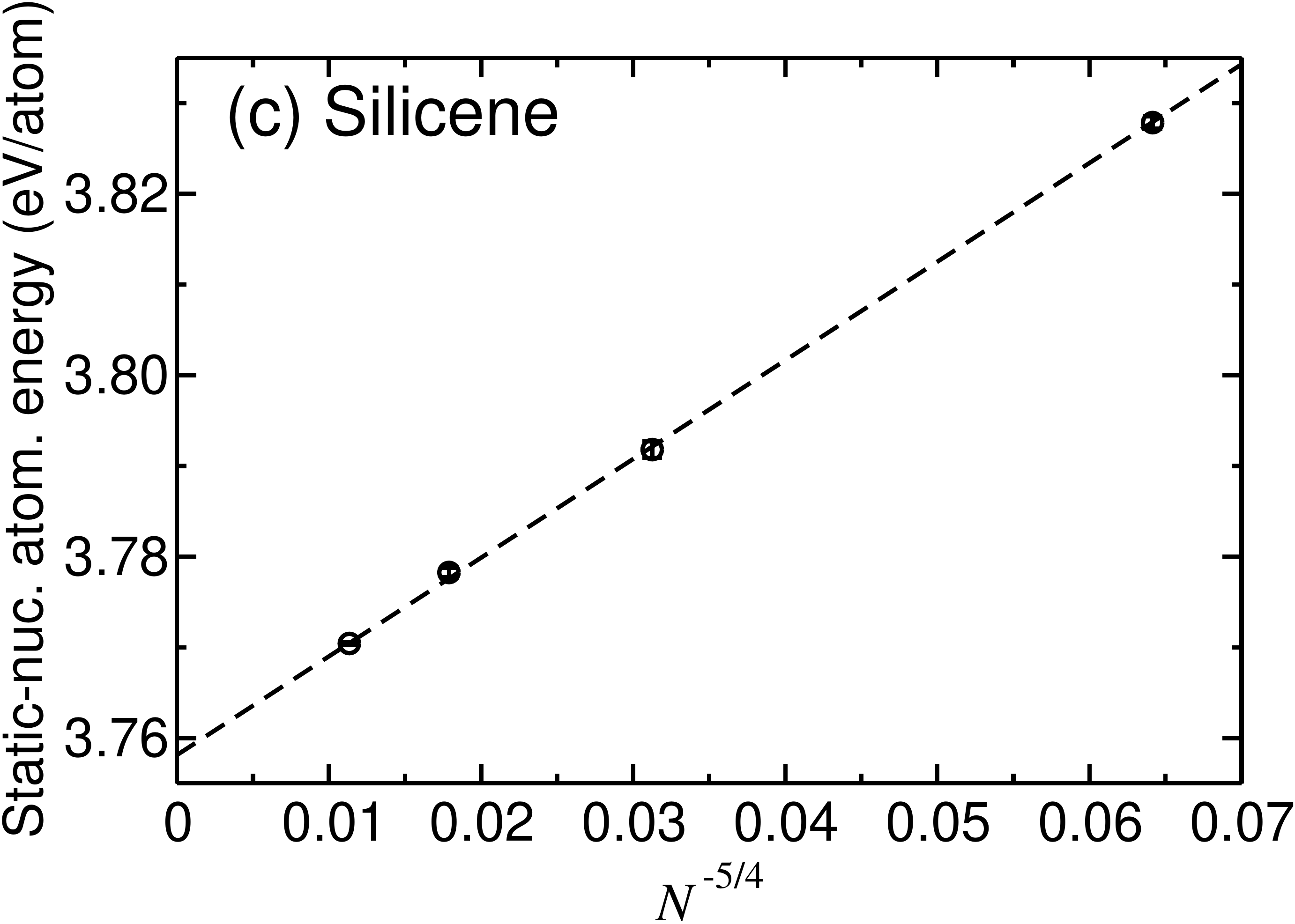}
\caption{TA-DMC static-nucleus atomization energies of (a) graphene,
(b) bulk silicon, and (c) silicene against $N^{-5/4}$ for graphene
and silicene, and $N^{-1}$ for bulk silicon, where $N$ is the number
of primitive cells in the supercell.
The atomization energies are defined with respect to the DMC
spin-polarized $^3\text{P}_0$ ground states of isolated carbon and
silicon atoms.}
\label{fig:atomization_energies}
\end{figure}

DMC atomization energies are plotted against system size in Fig.\
\ref{fig:atomization_energies} for graphene, bulk silicon, and
silicene, respectively, showing that finite-size effects are largely
removed by extrapolation.
DFT-PBE vibrational Helmholtz free energies are reported in Table
\ref{tab:vib_en_pristine}.
In graphene the vibrational free energy is relatively small, due to
the extreme stiffness of the lattice.
At room temperature, vibrational effects stabilize silicene with
respect to bulk diamond-structure silicon.

\begin{table}[!htpb]
\caption{DFT-PBE vibrational Helmholtz free energies per atom for
  pristine graphene, bulk silicon, and silicene.}
\label{tab:vib_en_pristine} \begin{ruledtabular}
\begin{tabular}{lccc}

\multirow{2}{*}{Temperature (K)} & \multicolumn{3}{c}{Vibrational free
  energy (meV/atom)} \\

& Graphene & Silicon & Silicene \\

\hline

0 & $3.0$ & $61.9$ & $50.6$ \\

298 & $3.2$ & $36.1$ & $3.0$ \\

\end{tabular}
\end{ruledtabular}
\end{table}

\begin{table*}[!htpb]
\caption{\label{tab:atomization_energy} Helmholtz free energies of
  atomization of graphene, bulk silicon, and silicene.
The DFT-PBE vibrational free-energies shown in Table
\ref{tab:vib_en_pristine} have been subtracted from our static-nucleus
atomization energies.
Unlike the present work, the silicene DFT calculations in
Refs.\ \onlinecite{cahangirov2009two} and
\onlinecite{drummond2012electrically} used a non-spin-polarized
($^1$S$_0$) Si atom as the atomic reference state, obtaining DFT-LDA
atomization energies of $5.06$ \cite{cahangirov2009two} and $5.12$ eV
\cite{drummond2012electrically} and DFT-PBE and DFT-HSE06 atomization
energies of $4.69$ and $4.70$ eV, respectively
\cite{drummond2012electrically}.
The difference between the single-determinant DMC energies of a
non-spin-polarized ($^1$S$_0$) and a spin-polarized ($^3\text{P}_0$)
silicon atom is $1.1506(6)$ eV\@.
The difference is $0.6107$ eV within DFT-LDA and $0.7987$ eV within
DFT-PBE\@.
The atomic reference state for the graphene atomization energies is
the spin-polarized ($^3\text{P}_0$) ground state of an isolated carbon
atom.
Results without citations were obtained in the present work.
}
\begin{ruledtabular}
\begin{tabular}{lcccccc}

\multirow{2}{*}{Method}& \multicolumn{6}{c}{Atomization energies
  (eV/atom)} \\

& \multicolumn{2}{c}{Graphene} & \multicolumn{2}{c}{Bulk silicon} &
\multicolumn{2}{c}{Silicene} \\

\hline

Temperature & $0$ K & $298$ K & $0$ K & $298$ K & $0$ K & $298$ K
\\

\hline

DFT-LDA & $8.96$ \cite{Graziano2012}, $8.632$ \cite{Mostaani_2015},
$8.912$ & $8.912$ & $5.34$ \cite{alfe2004diamond}, $5.3$
\cite{li1991cohesive}, $5.29$ & $5.31$ & $4.5414$ & $4.5869$ \\

DFT-PW91 & & & $4.653$ \cite{alfe2004diamond} & & \\

DFT-PBE & $7.93$ \cite{Graziano2012}, $7.873$ \cite{Mostaani_2015},
$7.916$ & $7.915$ & $4.55$ & $4.58$ & $3.9238$ & $3.9714$ \\

GFMC & & & $4.51(3)$ \cite{li1991cohesive} \footnote{Green's function
  Monte Carlo method.} & \\

DMC & $7.395(3)$ \cite{Mostaani_2015}, $7.388(2)$ & $7.388(2)$ & $4.62(1)$ \cite{alfe2004diamond}, $4.63(2)$
\cite{leung1999calculations}, $4.4815(6)$ &$4.5073(6)$ & $3.7075(4)$ &
$3.7551(4)$ \\

Experiment & $7.357(5)$ \cite{Haynes2010} & & $4.62(8)$
\cite{farid1991cohesive,NIST_Si_atomization} & & \\

\end{tabular}
\end{ruledtabular}
\end{table*}

Vibrationally corrected DMC atomization energies for free-standing
graphene, bulk silicon, and free-standing silicene, extrapolated to
infinite system size, are reported in Table
\ref{tab:atomization_energy}, along with DFT results.
The atomization energies of both bulk silicon and silicene are
significantly overestimated in DFT compared with DMC\@.
Bulk silicon is energetically more stable than silicene by a huge
margin of $0.7522(5)$ eV/atom.
By contrast, the atomization energies of graphite, graphene, and
carbon diamond are very similar, at about $7.43$ eV/atom
\cite{CRC,Mostaani_2015}.
Our DMC result for graphene compares extremely well with the
experimental result for graphite \cite{Haynes2010}, the two differing
by only $\sim 0.02$ eV/atom.
For the atomization energy of bulk silicon there is a small but
statistically significant difference between our DMC result and
earlier works \cite{alfe2004diamond,leung1999calculations}; this is
probably due to the fact that the earlier works did not use twist
averaging.
The DMC and DFT-PBE results are in reasonable agreement with the
experimentally determined atomization energy of bulk silicon.

% }}*

% *{{ Conclusions

\section{Conclusions}
\label{sec:conc}

We have used QMC methods to investigate the accuracy of DFT in
first-principles studies of point defect formation in monolayer
graphene.
Over accessible ranges of supercell size ($3 \times 3$--$5\times 5$
primitive cells), both DFT and QMC formation energies are affected by
both systematic and quasirandom finite-concentration effects on a
$\sim 1$ eV energy scale.
Systematic finite-concentration effects are similar in QMC and DFT,
but the difference between the QMC and DFT formation energies is still
subject to quasirandom errors, of order $0.5$ eV\@.
To reduce these errors, the difference between QMC and DFT formation
energies may be averaged over supercell sizes, providing a correction
that can be applied to the DFT formation energy obtained using large
supercell sizes.
We find that DFT-PBE underestimates the formation energies of isolated
monovacancies, silicon substitutions, and Stone-Wales defects by a
significant margin of order 1 eV\@.
Vibrational contributions to the free energies of formation of point
defects in graphene have also been found to be non-negligible, on a
0.5--1 eV energy scale.
Thus there are many factors to balance when evaluating defect
formation energies in 2D materials from first principles.
Similarly challenging behavior is expected for related quantities such
as defect migration energy barriers.

We have also compared the QMC atomization energies of monolayer
graphene, silicene, and bulk silicon, finding that bulk silicon is
more stable than silicene by $0.7522(5)$ eV per atom.
This quantifies the significant thermodynamic challenge involved in
producing free-standing silicene.

% }}*

% *{{ Acknowledgements
\begin{acknowledgments}
D.M.T.\ is fully funded by the Graphene NOWNANO CDT (EPSRC Grant
No.\ EP/L01548X/1).
This work was performed using resources provided by the Cambridge
Service for Data Driven Discovery (CSD3) operated by the University of
Cambridge Research Computing Service (\url{www.csd3.cam.ac.uk}),
provided by Dell EMC and Intel using Tier-2 funding from the
Engineering and Physical Sciences Research Council (capital grant
EP/P020259/1), and DiRAC funding from the Science and Technology
Facilities Council (\url{www.dirac.ac.uk}).
Additional computer resources were provided by Lancaster University's
High End Computing cluster.
\end{acknowledgments}

% }}*

% *{{ bibliography setup
\bibliography{graphene}
% }}*

% *{{ Appendix

\appendix

\section{Computational details}
\label{app:comp_details}

\subsection{DFT calculations}

\subsubsection{Total energy, geometry optimization, and phonon calculations}

Our DFT calculations were performed using the PBE generalized gradient
approximation exchange-correlation functional \cite{Perdew_1996} and
the plane-wave-basis code \textsc{castep} \cite{Clark2005}.
The total energy, geometry optimization, and phonon calculations all
used ultrasoft pseudopotentials \cite{Vanderbilt1990} to represent the
nuclei and core electrons.
A plane-wave cutoff energy of $556$ eV was used for pristine and
defective graphene and a cutoff energy of $305$ eV was used for bulk
silicon and silicene.
For pristine graphene and silicene, the total energies were calculated
using $51\times51$ and $53 \times 53$ Monkhorst-Pack ${\bf k}$-point
grids, respectively.
The total energies of defective graphene were calculated for
supercells of $N$ primitive cells in a $\sqrt{N}\times \sqrt{N}$
arrangement containing a single defect, using Monkhorst-Pack grids of
approximately $(51/\sqrt{N}) \times (51/\sqrt{N})$ ${\bf k}$-points;
in the following we refer to the use of these grids as ``fine'' ${\bf
  k}$-point sampling.
The geometry in each of the defective graphene supercells was
optimized to a force tolerance of $0.0025$ eV\,{\AA}$^{-1}$ with fixed
lattice vectors corresponding to a pristine-graphene carbon-carbon
bond length of $1.42$ {\AA} \cite{Kelly1981,Dresselhaus1988}.
Our bulk silicon calculations used $17\times17\times17$ Monkhorst-Pack
${\bf k}$-point grids.
All our 2D DFT calculations were performed using an artificial
periodicity of 30 bohr in the out-of-plane direction.
Non-spin-polarized calculations were used except where stated
otherwise.

Phonon calculations using the finite displacement method in DFT were
used to evaluate the vibrational contributions to the free energy.
These calculations were performed using atomic displacements of
$0.005$, $0.01$, $0.015$, $0.02$, and $0.025$ bohr, with the final
energies obtained by linearly extrapolating to zero atomic
displacement.
For each supercell $5\times5$ Monkhorst-Pack supercell ${\bf k}$-point
grids were used.
Geometries were first optimized to a force tolerance of $0.0005$
eV\,{\AA}$^{-1}$.

\subsubsection{QMC orbital generation}
\label{sec:orbitals}

Our DFT orbital-generation calculations used the PBE functional
together with Trail-Needs Dirac-Fock pseudopotentials
\cite{Trail2005a,Trail2005b} to represent the nuclei and core
electrons, with $s$ being the angular momentum of the local component
when the pseudopotentials are re-represented in Kleinman-Bylander
form \cite{Kleinman_1982}.
The geometry was fixed at the DFT-PBE geometry obtained using
ultrasoft pseudopotentials.
The graphene supercells used for the QMC calculations consisted of
$3\times3$, $4\times4$, and $5\times5$ primitive cells, where the
plane-wave cutoff energy for the smaller two supercells was $3401$ eV,
and the plane-wave cutoff energy for the larger supercell was $2231$
eV\@.
These cutoff energies are such that the DFT energy per atom is
converged to within, respectively, $0.1$ mHa and $1.59$ mHa (known as
chemical accuracy) \cite{drummond2016trail}.
For bulk silicon, supercells of $2\times2\times2$, $3\times3\times3$,
and $4\times4\times4$ primitive cells were used with a plane-wave cutoff
energy of $2231$ eV for all system sizes, while the silicene supercells
comprised $3\times3$ and $6\times6$ arrays of primitive cells.
An artificial periodicity of $30$ bohr was used for the graphene and
silicene calculations.
Non-spin-polarized DFT calculations were used except where otherwise
stated.

\subsection{QMC calculations}

\subsubsection{Trial wave functions}

The trial wave functions used for the QMC calculations were of
Slater-Jastrow (SJ) form, containing a product of determinants of
spin-up and spin-down orbitals; see Sec.\ \ref{sec:orbitals}.
Different sets of orbitals were generated for each twist (i.e., offset
${\bf k}_\text{s}$ to the grid of Bloch ${\bf k}$ vectors).
The plane-wave orbitals were re-represented in a blip (B-spline) basis
\cite{Alfe2004} both for computational efficiency in the QMC
calculations and to remove the unwanted periodicity in the
out-of-plane direction.
The Jastrow factor, a nodeless function of the interparticle distances
containing optimizable free parameters, consisted of polynomial
electron-electron, electron-nucleus, and electron-electron-nucleus
terms, and plane-wave electron-electron terms \cite{Drummond2004}.
Trial wave functions were optimized first by minimizing the variance
of the energy \cite{Umrigar1988,Drummond2005} and then by minimizing
the energy expectation value \cite{Umrigar2007}.
For a given supercell, this optimization was performed at a single,
randomly chosen twist, with the resulting Jastrow factor being used at
all twists.

For some test cases at individual twists, Slater-Jastrow-backflow
trial wave functions were used to investigate the fixed-node errors in
our SJ-DMC results.
These wave functions were obtained by optimizing the backflow and
Jastrow parameters together using energy minimization.
The backflow functions contained polynomial electron-electron and
electron-nucleus terms \cite{Lopez_2006}.
Further tests using a long-range plane-wave electron-electron backflow
function were also carried out: see Sec.\ \ref{sec:backflow}.

Trail-Needs Dirac-Fock pseudopotentials \cite{Trail2005a,Trail2005b}
were used to represent the ionic cores, with $d$ being the angular
momentum of the local component.

\subsubsection{DMC calculations}

\begin{figure}[!htpb]
\centering
\includegraphics[scale=0.28,clip]{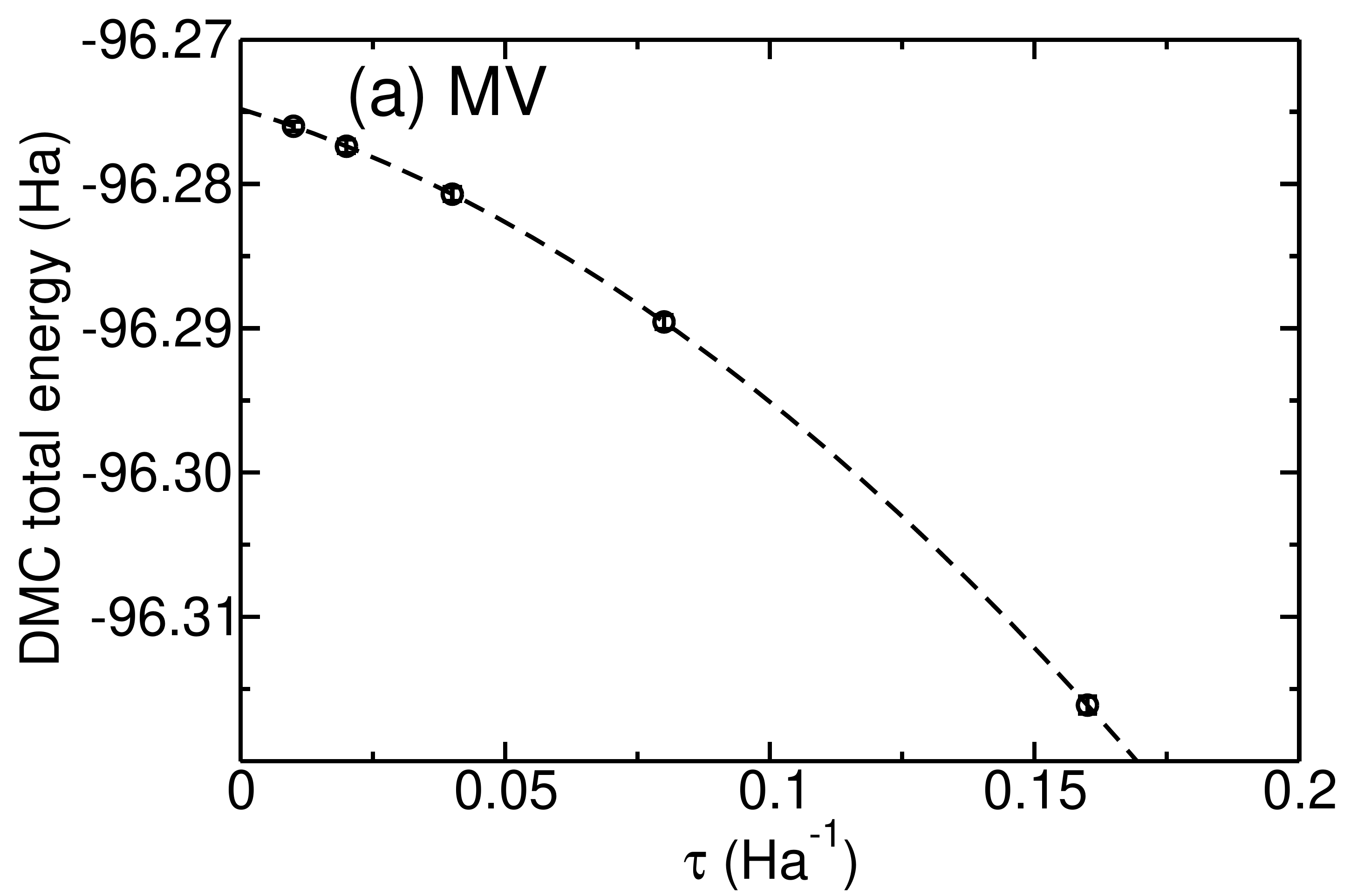}
\\ \includegraphics[scale=0.28,clip]{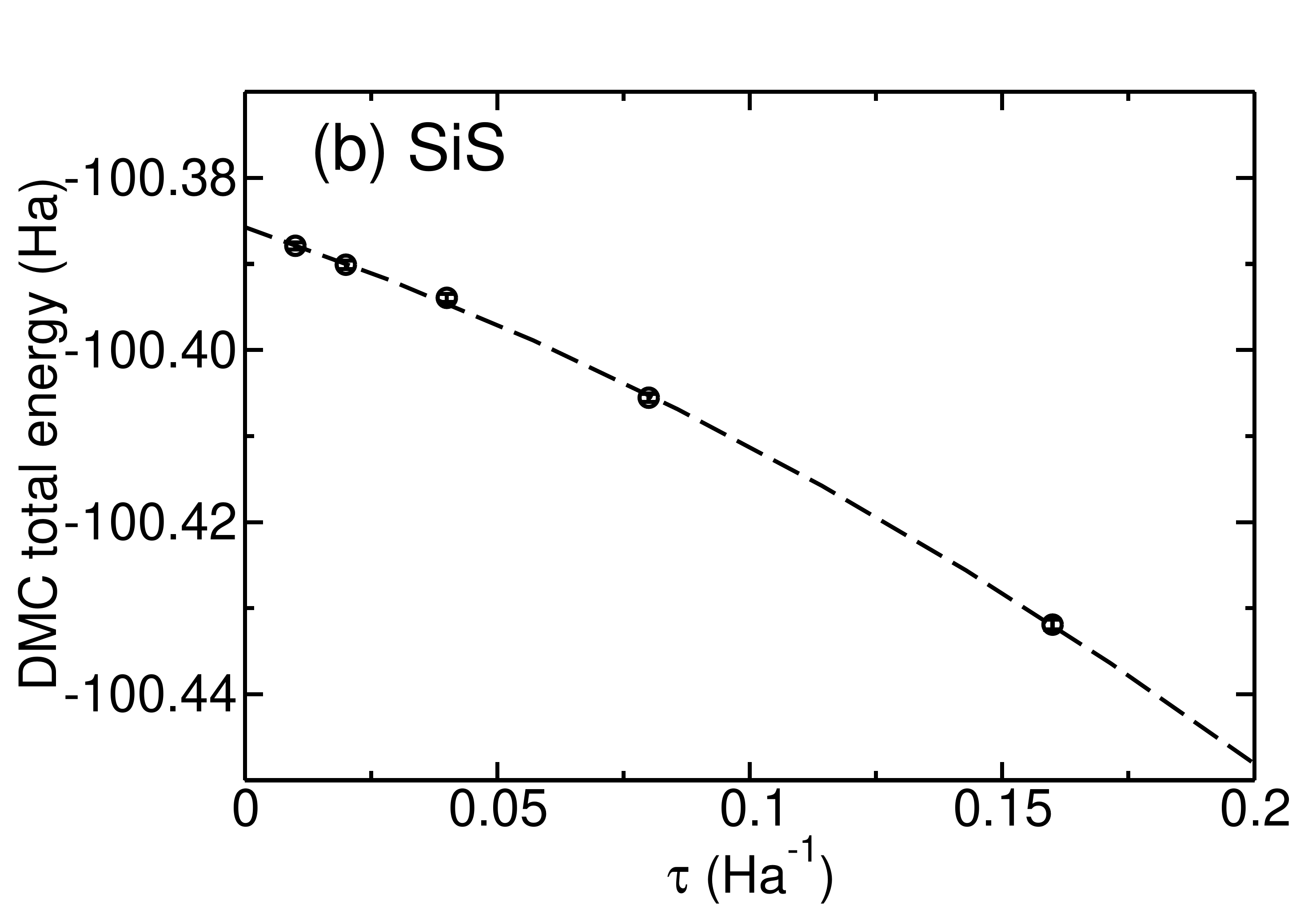}
\\ \includegraphics[scale=0.28,clip]{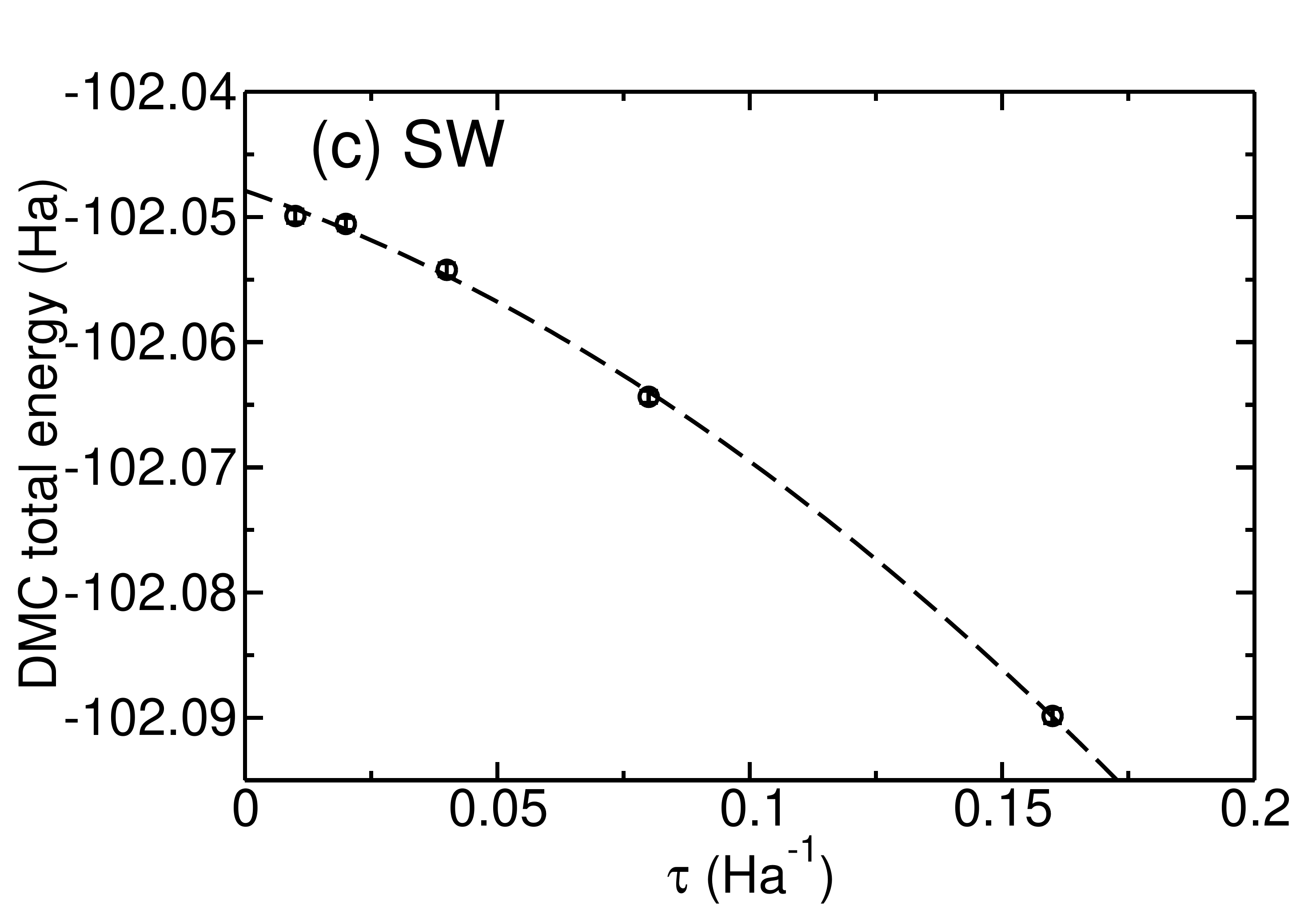}
\caption{DMC total energies per supercell of (a) MV, (b) SiS, and (c)
  SW defects in a $3\times3$ supercell of graphene against DMC time
  step $\tau$ at a single, randomly chosen twist
  $\mathbf{k}_\text{s}$.
The dashed lines show quadratic fits to the energy as a function of
time step.
\label{fig:total_tsb_individual_twist}}
\end{figure}

\begin{figure}[!htpb]
\centering
\includegraphics[scale=0.28,clip]{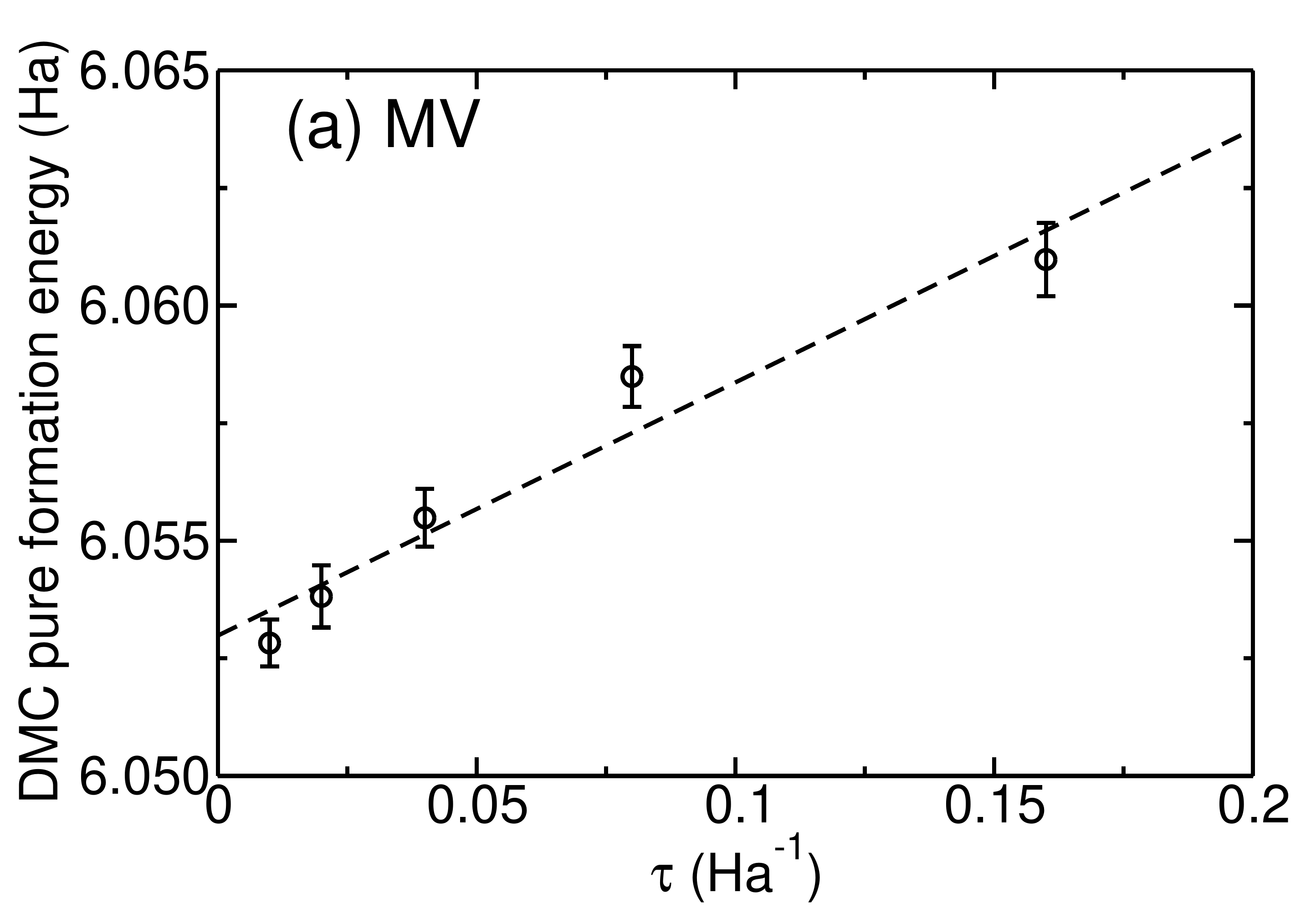}
\\ \includegraphics[scale=0.28,clip]{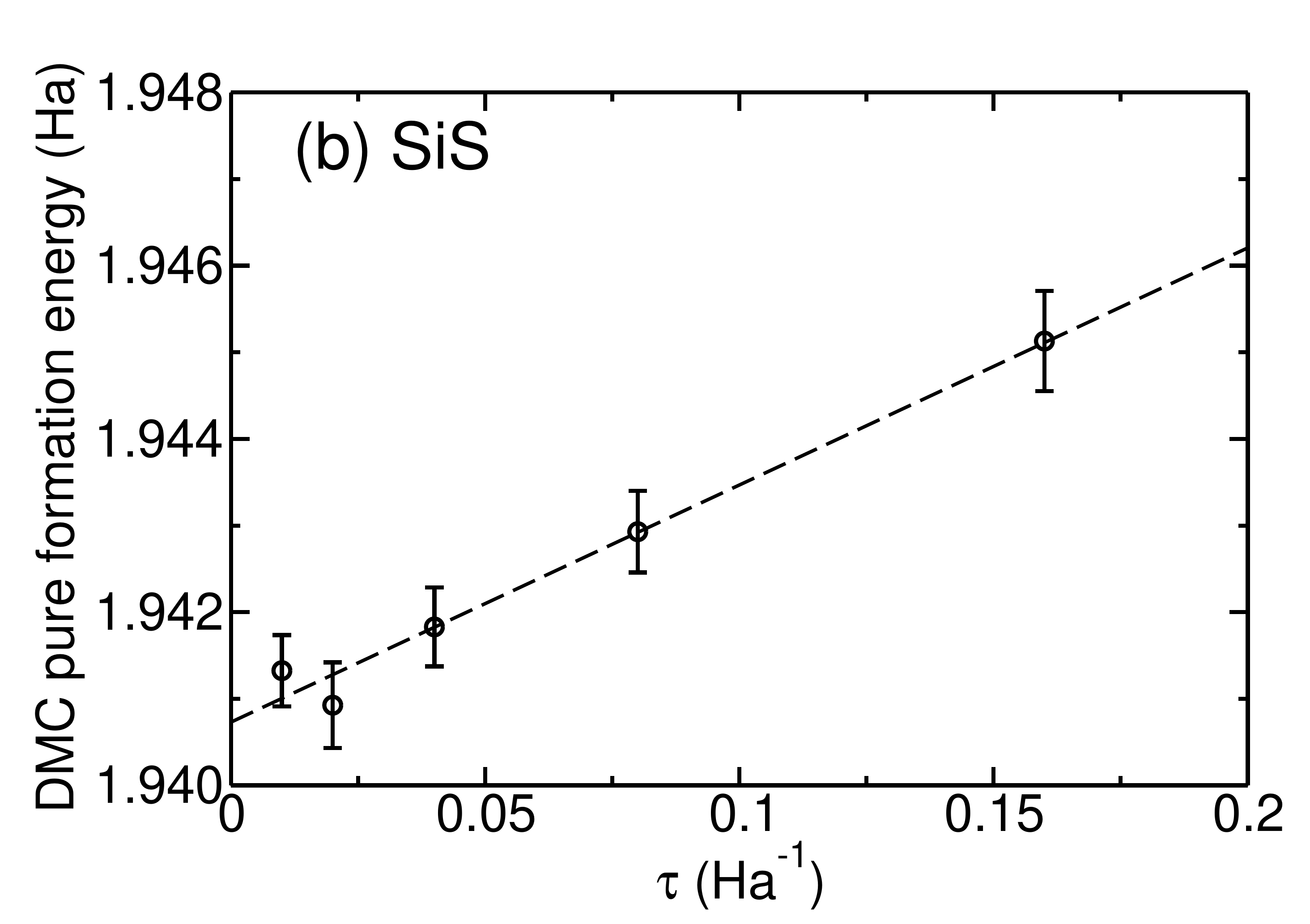}
\\ \includegraphics[scale=0.28,clip]{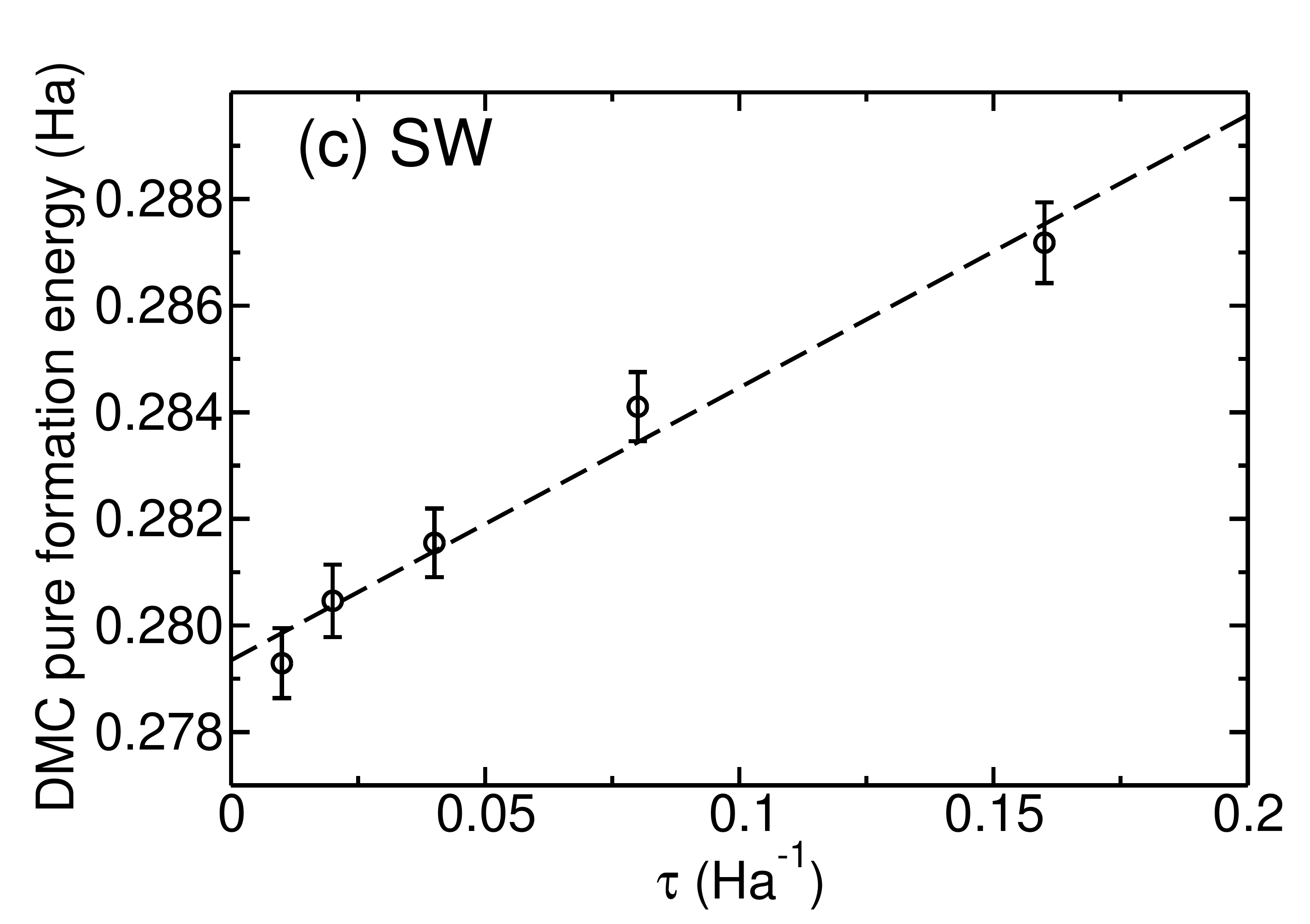}
\caption{DMC pure formation energies of (a) MV, (b) SiS, and (c) SW
  defects in a $3\times3$ supercell of graphene against DMC time step
  $\tau$ at the twist $\mathbf{k}_\text{s}$ used in
  Fig.\ \ref{fig:total_tsb_individual_twist}.
The dashed lines show linear fits to the pure formation energy as a
function of time step.
\label{fig:pure_tsb_individual_twist}}
\end{figure}

To calculate the pure defect formation energy of each of the three
defects we have studied in graphene, pairs of DMC calculations were
carried out at each twist in all the defective and pristine graphene
supercells.
Time steps of $\tau=0.04$ and $0.16$ Ha$^{-1}$ were used in these
calculations, with the corresponding target walker populations being
varied in inverse proportion to the time step.
In all cases the target population was at least 256 walkers.
The energies were then extrapolated linearly to zero time step.
For the total energies of defective and pristine graphene we would not
expect these time steps to be small enough to be in the linear bias
regime (as confirmed by the results shown in
Fig.\ \ref{fig:total_tsb_individual_twist}); however, as shown in
Fig.\ \ref{fig:pure_tsb_individual_twist}, the nonlinear parts of the
time-step bias largely cancel out of the pure defect formation energy.

To calculate the energies per atom of graphene and bulk silicon we
used smaller time steps of $\tau=0.01$ and $0.04$ Ha$^{-1}$, allowing
time-step bias in the total energy per atom to be largely removed by
linear extrapolation.
Again, we varied the target walker population inversely with time
step.

% }}*

\end{document}